\documentclass[11pt]{article}
\pdfoutput=1
\usepackage{amsmath}
\usepackage{amssymb}
\usepackage{graphicx,bbm,mathrsfs}
\usepackage{nicefrac}
\usepackage{slashed}
\usepackage{bbm}
\usepackage{geometry}
\usepackage{empheq}
\usepackage{stackrel}
\usepackage{ulem}

\usepackage{jheppub}

\usepackage{mathtools}
\usepackage{enumitem}
\usepackage{dcolumn}   
\usepackage{bm}        
\usepackage{graphicx,mathrsfs}
\usepackage{nicefrac}
\usepackage{multirow}
\usepackage{color}
\usepackage{mathtools}
\usepackage{makecell}

\usepackage{environ} 
\usepackage{lipsum} 
 \NewEnviron{Smaller11}{
           \scalebox{1.1}{$\BODY$} 
 } 
 \NewEnviron{Smaller08}{
           \scalebox{0.8}{$\BODY$} 
 }

\newcommand{\be}{\begin{equation}}
\newcommand{\ee}{\end{equation}}
\newcommand{\bea}{\begin{eqnarray}}
\newcommand{\eea}{\end{eqnarray}}

\newcommand{\x}{\mathbf{x}}

\newcommand{\AdSS}{\textrm{\tiny AdSS}}
\newcommand{\LD}{\textrm{\tiny LD}}
\newcommand{\ALD}{\textrm{\tiny ALD}}

\newcommand{\brane}{b}

\usepackage{wasysym}
\usepackage{hhline,colortbl}

\newcommand{\nn}{\nonumber}

  \newcommand{\step}{\xi}
  \newcommand{\gap}{\sigma}
  
  \newcommand{\tscale}{\tilde\Lambda}

\usepackage{titlesec}

\titleformat*{\section}{\Large\bfseries}
\titleformat*{\subsection}{\large\bfseries}
\titleformat*{\subsubsection}{\large\bfseries}
\titleformat*{\paragraph}{\large\bfseries}
\titleformat*{\subparagraph}{\large\bfseries}

\makeatletter
\newcommand*{\prodsym}{%
  \DOTSB
  \mathop{
    \mathchoice
      {\rlap{\kern.3em\rotatebox[origin=c]{-90}{}}{\prod}}
      {\vcenter{\rlap{\kern.2em\rotatebox[origin=c]{-90}{}}}{\prod}}
      {\sum}{\sum}
  }\slimits@
}
\makeatother

\makeatletter
\DeclareFontFamily{OMX}{MnSymbolE}{}
\DeclareSymbolFont{MnLargeSymbols}{OMX}{MnSymbolE}{m}{n}
\SetSymbolFont{MnLargeSymbols}{bold}{OMX}{MnSymbolE}{b}{n}
\DeclareFontShape{OMX}{MnSymbolE}{m}{n}{
    <-6>  MnSymbolE5
   <6-7>  MnSymbolE6
   <7-8>  MnSymbolE7
   <8-9>  MnSymbolE8
   <9-10> MnSymbolE9
  <10-12> MnSymbolE10
  <12->   MnSymbolE12
}{}
\DeclareFontShape{OMX}{MnSymbolE}{b}{n}{
    <-6>  MnSymbolE-Bold5
   <6-7>  MnSymbolE-Bold6
   <7-8>  MnSymbolE-Bold7
   <8-9>  MnSymbolE-Bold8
   <9-10> MnSymbolE-Bold9
  <10-12> MnSymbolE-Bold10
  <12->   MnSymbolE-Bold12
}{}

\let\llangle\@undefined
\let\rrangle\@undefined
\DeclareMathDelimiter{\llangle}{\mathopen}%
                     {MnLargeSymbols}{'164}{MnLargeSymbols}{'164}
\DeclareMathDelimiter{\rrangle}{\mathclose}%
                     {MnLargeSymbols}{'171}{MnLargeSymbols}{'171}
\makeatother

\begin{document}

\vspace*{4mm}

\begin{center}

\thispagestyle{empty}
{
\LARGE\sc
 On Continuum Effective Field Theories, 
  \\ \vspace{0.5cm}  Gravity and Holography
}\\[12mm]

\renewcommand{\thefootnote}{\fnsymbol{footnote}}

{\large  
Sylvain~Fichet$^{\,a,b}$ \footnote{sfichet@caltech.edu }\,, 
Eugenio~Meg\'{\i}as$^{\, c}$ \footnote{emegias@ugr.es}\,,
Mariano~Quir\'os$^{\, d}$ \footnote{quiros@ifae.es}\,
}\\[12mm]
\end{center} 
\noindent

${}^a\!$ 
\textit{ICTP South American Institute for Fundamental Research  \& IFT-UNESP,} \\
\indent \, \textit{R. Dr. Bento Teobaldo Ferraz 271, S\~ao Paulo, Brazil}

${}^b\!$ 
\textit{Centro de Ciencias Naturais e Humanas, Universidade Federal do ABC,} \\
\indent \; \textit{Santo Andre, 09210-580 SP, Brazil}

${}^c\!$ 
\textit{Departamento de F\'{\i}sica At\'omica, Molecular y Nuclear and} \\
\indent \; \textit{Instituto Carlos I de F\'{\i}sica Te\'orica y Computacional,} \\
\indent \; \textit{Universidad de Granada, Avenida de Fuente Nueva s/n, 18071 Granada, Spain}

${}^d\!$  
\textit{Institut de F\'{\i}sica d'Altes Energies (IFAE) and} \\
\indent \; \textit{The Barcelona Institute of  Science and Technology (BIST),} \\
\indent \; \textit{Campus UAB, 08193 Bellaterra, Barcelona, Spain}

\addtocounter{footnote}{-1}

\vspace*{15mm}

\begin{center}
{  \bf  Abstract }
\end{center}

We examine effective field theories (EFTs) with a continuum sector in the presence of gravity. We first explain, via  arguments  based on central charge and species scale, that an EFT with a free continuum cannot consistently couple to standard (\textit{i.e.}~4D~Einstein)  gravity. It follows that EFTs with a free or nearly-free continuum must either have a finite number of degrees of freedom or nonstandard gravity. 
The latter claim  is realized for holographically-defined continuum models. 
We  demonstrate  this by computing the deviations from standard gravity 
in a specific 5D scalar-gravity system  that gives rise to a gapped continuum (\textit{i.e.}~the linear dilaton background). We  find an $R^{-2}$ deviation from the Newtonian potential.  
At finite temperature we find an energy density with matter-like behavior in the brane Friedmann equation,  holographically induced from the bulk geometry. Thus, remarkably, a brane-world living in the linear dilaton background automatically contains dark matter.  
We also present a slightly more evolved asymptotically--AdS linear dilaton model, for which the deviations exhibit 
a transition between AdS and linear dilaton behaviors.

\newpage
\setcounter{tocdepth}{2}
\tableofcontents

\newpage

\section{Introduction}
\label{se:intro}

Among the multitude of  effective field theories (EFT) extending the Standard Model (SM) of particles physics, models involving a continuum sector stand out as an intriguing possibility. Of course, any weakly coupled Poincaré-invariant quantum field theory (QFT) features a continuum in its spectral distributions. But beyond this standard case, a nearly-free continuum can also emerge in theories with some nontrivial underlying dynamics.
 Such a continuum can for example appear from gauge sectors with large number of colors, or from EFTs involving a brane (\textit{i.e.}~domain wall or defect) living in a higher dimensional spacetime. Such nontrivial continuum sectors are the subject of this work. 

From a more phenomenological viewpoint we may also write, from a bottom-up approach, a continuum model with arbitrary spectral functions describing the phenomena that are potentially observables in a given set of experiments, as  allowed by the  rules of the EFT paradigm.  
Phenomenologically, the underlying dynamics of the continuum may, or may not, matter, depending on the  situation. In certain cases, it may be sufficient to use an effective model in which the continuum has properties analogous to an ordinary free field. This is called a generalized free field~\cite{Greenberg:1961mr}. This approach applies for example to processes observable at colliders, such as ``SM $\to$ continuum $\to$ SM'' and ``SM $\to$ continuum'' for which only the two-point function of the continuum is needed. 
Regarding the latter class of processes, we emphasize that even though a continuum does not have well defined asymptotic states, such processes make sense as inclusive ones, for which no measurement of the continuum final state is required.

The EFT of a free continuum works fine for scattering processes observable at a collider. But, are there other physical observables for which the description of a continuum as a generalized free field does not apply? The answer is positive:  whenever interactions with gravity are considered, the underlying dynamics of the continuum does matter. Clarifying the interplay of continuum models with gravity is the first aim of this work. This investigation will then naturally lead us to explore aspects of gravity in holographic models of continuum, which is its second aim.\,\footnote{Along this line,  the companion paper~\cite{DM_LD} focuses  on the emergence of cosmological dark matter  in the linear dilaton background.} 
 Our analysis is structured as follows. 
 
 The first part of the paper contains a broad  analysis of continuum EFTs.
In Sec.~\ref{se:models} we lay out the formalism, and introduce the notion of generalized free field. In Sec.~\ref{se:gravity} we review the arguments (both old and new) that prevent a generalized free field to consistently couple to standard gravity. As an interesting aside we give an argument for the species scale that is valid for conformal field theories (CFTs) with any central charge and coupling. We then discuss the classes of models that give rise to a gravity-compatible continuum. It turns out that, apart from the conformal case, continuum models are best studied via 5-dimensional holographic models. 

The second part of the paper is focussed on gravity and cosmology of  holographically-defined continuum models.
In Sec.~\ref{se:holcontinuum} we lay out the basic holographic framework, review necessary QFT aspects and show how to compute the deviations from the standard Friedmann equation and Newtonian potential. 
In Sec.~\ref{se:lineardilaton} we solve two versions of a specific scalar-gravity background (both analytically and numerically) that features a gapped continuum and investigate gravity aspects.   Section~\ref{se:conclusion} contains a summary. 
Finally App.~\ref{app:discretum} contains further discussions on the transition between discretum and continuum, App.~\ref{app:Technical_details} includes
technical details on the solutions of the 5D scalar-gravity system, and App.\,\ref{app:5D_continuity_eq} explicitly computes the  conservation law in the linear dilaton and asymptotically AdS linear dilaton backgrounds.

\paragraph{Previous literature:} 
The phenomenological possibility of continuum models was first highlighted in Refs.~\cite{Georgi:2007ek,Georgi:2007si}.   We here mention only a few of the   subsequent developments as an introduction into the literature, Refs.~\,\cite{Stephanov:2007ry,Strassler:2008bv,Cacciapaglia:2008ns,Friedland:2009iy,Friedland:2009zg}. %
Aspects of cosmology with a conformal sector have been investigated in, \textit{e.g.}~Refs.~\cite{vonHarling:2008vwq,vonHarling:2012sz,Hong:2019nwd,Redi:2020ffc,Hong:2022gzo}, and in Refs.~\cite{Grzadkowski:2008xi,Artymowski:2019cdg,Artymowski:2021fkw}
in case of large $N$ weakly coupled CFT. 
A continuum as a mediator in the dark sector has been investigated in Refs.~\cite{Katz:2015zba,Chaffey:2021tmj}. Finally
a proposal of ``continuum dark matter'' was recently made in Refs.~\cite{Csaki:2021gfm,Csaki:2021xpy}, that can be put in perspective with the arguments and results of the present work.

\paragraph{Conventions:}

Throughout this work we use the conventions of Misner-Thorne-Wheeler  \cite{Misner:1973prb}, which include the mostly-plus  metric signature ${\rm sgn}(g_{\mu\nu})=(-,+,\ldots,+)$.
Likewise we  define $\sqrt{g}\equiv \sqrt{|\det{g_{\mu\nu}}|}$ for an arbitrary metric $g_{\mu\nu}$ in  any number of spacetime dimensions. 
In momentum space, a particle with mass $m$ on the mass shell satisfies $-p^2=m^2$, where $p^2=p^\mu p^\nu g_{\mu\nu}$. 
When discussing quantum fields,  we will sometimes introduce the opposite  quantity $q^2\equiv-p^2$, such that the correlators expressed in terms of $q^2$  match the formulas that would be obtained with the mostly-minus metric, which is a more common convention for particle physics.

\section{ Continuum models}
\label{se:models}
In this section we discuss some basic aspects of continuum models and introduce the notion of free continuum limit. 
\subsection{Continuum EFT}

We consider an EFT described by the following four-dimensional Lagrangian, 
\be{\cal L}={\cal L}_{\rm particles}[\varphi] +  {\cal L}_{\rm continuum}[\Phi] + b\, \tilde {\cal O}[\varphi]{\cal O}[\Phi] \,.
\label{eq:L_eff}
\ee
This Lagrangian contains in general irrelevant operators~\footnote{The $\tilde {\cal O} {\cal O} $ operator is often taken as an irrelevant operator \textit{i.e.}~dim$[b]<0$. A detailed parametrization is unnecessary for our purposes.}. The fundamental fields $\varphi,\,\Phi$ can in principle have any spin. $\cal O$ and $\tilde {\cal O}$ are in general composite operators made of the corresponding fundamental fields. For simplicity the latter are assumed to be scalars.

The $\Phi$ sector is assumed to feature a nontrivial continuum --- in a sense defined further below and in Sec.~\ref{se:GFT}. 
 In the $\varphi$ sector, the spectral functions are assumed to describe stable or narrow particles as occurs in weakly coupled QFT. 
 The two sectors interact with each other only via the $\tilde {\cal O}[\varphi]{\cal O}[\Phi]$ operator.
From the viewpoint of an observer able to probe the particle sector ${\cal L}_{\rm particle}[\varphi]$,
the continuum sector is probed by the $\varphi$ fields through the $\tilde {\cal O} {\cal O} $ operator. Thus in the correlators of the $\varphi$ fields, the continuum sector manifests itself via subdiagrams made out of the correlators of ${\cal O}[\Phi]$, \textit{i.e.}~$\langle{\cal O}(x_1){\cal O}(x_2)\rangle$, $\langle{\cal O}(x_1){\cal O}(x_2){\cal O}(x_3)\rangle,\ldots$~\footnote{Time-ordering is left implicit, we use the usual shortcut notation $\left\langle {\cal O}(x) {\cal O}(0) \right\rangle=\left \langle \Omega | T\{ {\cal O}(x) {\cal O}(0) \} | \Omega \right\rangle$. }. Our interest precisely lies in these correlators of the continuum sector.

For any two-point (2pt) correlator  one can always introduce a spectral representation of the form\,\footnote{This follows from Cauchy's integral formula $f(a) = \frac{1}{2\pi i} \oint dz \frac{f(z)}{z-a}$.}\,
{
\be
\left\langle {\cal O}(x) {\cal O}(0) \right\rangle =
\int \frac{d^4p}{(2\pi)^4} e^{ip \cdot x} \int_{C} ds  \frac{i\,\rho(s)}{-p^2-s+i\epsilon}
\,\label{eq:spec}
\ee
}
where $\rho(s)$ is the spectral distribution and the contour $C$ encloses non-analyticites of the  correlator in momentum space. In our conventions the momentum is timelike for $p^2<0$. The  non-analyticities can either be poles or branch cuts along $\mathbb R_-$. 
On the domain corresponding to a branch cut the spectral density $\rho(s)$ is a smooth function.  In this most generic case we refer to the $\Phi$ sector as a \textit{continuum}.  As a more particular case, it may happen that the support of $\rho(s)$ be a discrete set of points. Similarly it is also possible that the function be  made of a series of narrow resonances such that the branch cut can be approximated by a set of points. 
 In such cases the spectral distribution describes a countable set of standard 4D particles, and we refer to the $\Phi$ sector more specifically as a \textit{discretum}.

\subsubsection{Interactions}

It is useful to classify the interactions encoded in the continuum sector.

\begin{enumerate}[label=\roman*)]
    \item  There are fundamental interactions between the $\Phi$ fields, encoded inside the Lagrangian ${\cal L}_{\rm continuum}$. We denote collectively these interactions by the coupling $g$. These fundamental interactions may be either weak or strong.

\item The continuum interacts with the particle sector via the $\tilde {\cal O}[\varphi]{\cal O}[\Phi]$ operator. This implies that local operators of the form
\be
{\cal L}_{\rm continuum}\supset g_{{\cal O},n} \left({\cal O}[\Phi](x)\right)^n
\ee
 are generically present in the continuum sector.  Analogous ones with an arbitrary number of derivatives also exist.  All these operators are in general present due to the quantum dynamics in the $\varphi$ sector.  In general, even if these operators are set to zero at a given scale, they are generated at a different scale due to renormalization group (RG) running. We refer collectively these local operators as ${\cal O}^n$  with corresponding coupling $g_{{\cal O},n}$. 
\end{enumerate}

\subsubsection{Realizations}
\label{se:realizations}

In principle \textit{any} interacting QFT  can realize the setup of Eq.~\eqref{eq:L_eff}. For example, for a weakly coupled interacting QFT the continuous part of the spectral distribution $\rho$ encodes  a multiparticle continuum, and possibly a resonance.
However, our interest lies in theories that can give rise to  a \textit{free continuum} when  some parametric limit is taken in the model (see Sec.~\ref{se:GFT}). A nontrivial dynamics is needed for such a limit to occur. It is realized in at least the two following classes of theories.

\begin{enumerate}[label=\it\roman*)]

\item  \textit{Gauge theories with a large number of colors $N$.} For simplicity we  assume that the $\Phi$ fields are in the adjoint representation, so that the standard large $N$ scaling applies~\cite{Witten:1979kh}. The theory may, in principle, have either weak or strong t'Hooft coupling $\lambda\equiv g^2N$. 
 As a particular case, the gauge theory may be at a conformal fixed point in which case it is a CFT. For $\lambda\ll 1$ this occurs at a Banks-Zaks fixed point if the theory has number of flavors within the conformal window. At $\lambda\ll1$  stringy effects are expected to emerge for large $N$ (see \textit{e.g.}~Refs.~\cite{PhysRevD.10.2445,PhysRevD.10.4262,tHooft:1977nqb,Nambu:1978bd,Luscher:1980iy,Luscher:1980ac,Sundrum:1997qt}) while at   $\lambda\gg1$  the stringy effects are expected to decouple \cite{Gubser:1998bc,Polchinski:2002jw}.

\item \textit{Holographic theories.} These arise from EFTs living in a 5D background (with  arbitrary metric)  featuring a flat 3-brane. 
In such a setup an effective Lagrangian of the form of Eq.\,\eqref{eq:L_eff} appears from the viewpoint of an observer placed on the brane. The $\varphi$ field is identified as a brane-localized mode with standard 4D spectral distribution, which mixes with a continuum controlled by the 5D dynamics (see Sec.~\ref{se:lineardilaton} for more details).
In such models there are both bulk and brane-localized local interactions, that we denote by $g_{\rm bulk}$ and $g_{\rm brane}$. We  also refer to this setup as a ``brane-world''  in the context of cosmological models. 

\end{enumerate}

\subsection{The free continuum (GFT) limit}
\label{se:GFT}

We are interested in taking a parametric limit for which a free continuum arises in the general Lagrangian  of Eq.\,\eqref{eq:L_eff}.   
Our notion of \textit{free continuum} is equivalent to the one described by a \textit{generalized free theory} (GFT), hence we are using either naming depending on context.   

GFTs have been studied in the context of  QFT and CFT (see~\textit{e.g.}~Refs.~\cite{Greenberg:1961mr,Dymarsky:2014zja,Kap:lecture}). 
In a GFT the connected part of the correlators of ${\cal O}$ vanishes. 
As a result the odd correlators are zero while the even correlators are given by the disconnected contributions  which are just a product of 2pt correlators.  For example for the 4pt correlator we have
\be \langle{\cal O}(x_1){\cal O}(x_2){\cal O}(x_3){\cal O}(x_4)\rangle=\langle{\cal O}(x_1){\cal O}(x_2)\rangle\langle{\cal O}(x_3){\cal O}(x_4)\rangle+{\rm permutations} \,.
\label{eq:4pt_free}
\ee  

We define the free continuum (\textit{i.e.}~GFT) limit, as the limit for which the fundamental interactions of the continuum sector vanish,
\be
{\cal L}_{\rm continuum}\bigg|_{g\to 0}\to {\cal L}_{\rm GFT}\, 
\label{eq:GFT_limit}
\ee
while the spectral density does \textit{not} become discrete (\textit{i.e.}~remains supported on $\mathbb{R}$ and not only on a discrete set of points when $g\to 0$).  
This definition of the free continuum limit automatically excludes the trivial case of an interacting QFT with finite degrees of freedom, since in that case for $g\to 0$ the multiparticle continuum vanishes and the spectral density of ${\cal O}[\Phi]$ becomes discrete. Thus some nontrivial dynamics in ${\cal L}_{\rm continuum}$ is required for a free continuum to emerge at $g\to 0$. 

Our definition of GFT allows for the existence of the local interactions ${\cal O}^n$. Thus in our definition the GFT correlators can have $O(g_{{\cal O},n})$ contributions.  This is however a minor point in the rest of our analysis, as we will obtain the same conclusions as if $g_{{\cal O},n}=0$.

How is the free continuum/GFT limit realized in the classes of models listed in the previous Sec.~\ref{se:realizations}? %
\begin{enumerate}[label=\it\roman*)]
\item
 A GFT emerges from  a gauge theory by taking the limit of infinite number of colors $N\to \infty$ at constant `t Hooft coupling --- notice that $g\to 0$ in this limit. 
Indeed, by normalizing the 2pt function coefficient such that it does not scale with $N$, standard large $N$ scaling arguments imply that the connected correlators scale as powers of $1/N$. In the $N\to\infty$ limit the odd correlators are $O(\frac{1}{N})$ and the even correlators are given by the free disconnected result plus $O(\frac{1}{N})$ terms. This matches the properties of a GFT, hence
\be
{\cal L}_{\rm gauge}\bigg|_{\begin{subarray}{l} N\to \infty\\
\lambda \, {\rm  fixed}  \end{subarray} }\to  {\cal L}_{\rm GFT} \,.
\ee
We could similarly write this limit for the full Lagrangian including the ${\cal O}^n$ interactions.

\item
 A GFT emerges from a holographic setup by sending the \textit{bulk couplings} to zero. Indeed in this limit the bulk propagators are free, hence the higher point correlators factorize into 2pt propagators.  The holographic theory inherits this property, therefore the holographic theory is a GFT.  
The brane couplings contribute solely to the ${\cal O}^n$ interactions --- which are allowed in our definition of GFT. 
In summary, for the full holographic Lagrangian we schematically have
\be
{\cal L}_{\rm hol}\bigg|_{g_{\rm bulk}\to 0} \to {\cal L}_{\rm particles}[\varphi] +  {\cal L}_{\rm GFT}[\Phi] + c \, \tilde {\cal O}[\varphi]{\cal O}[\Phi] \,.
\ee
\end{enumerate}

\begin{figure}[t]
    \centering
    \includegraphics[width=0.8\linewidth,trim={4cm 2.5cm 6cm 2.5cm},clip]{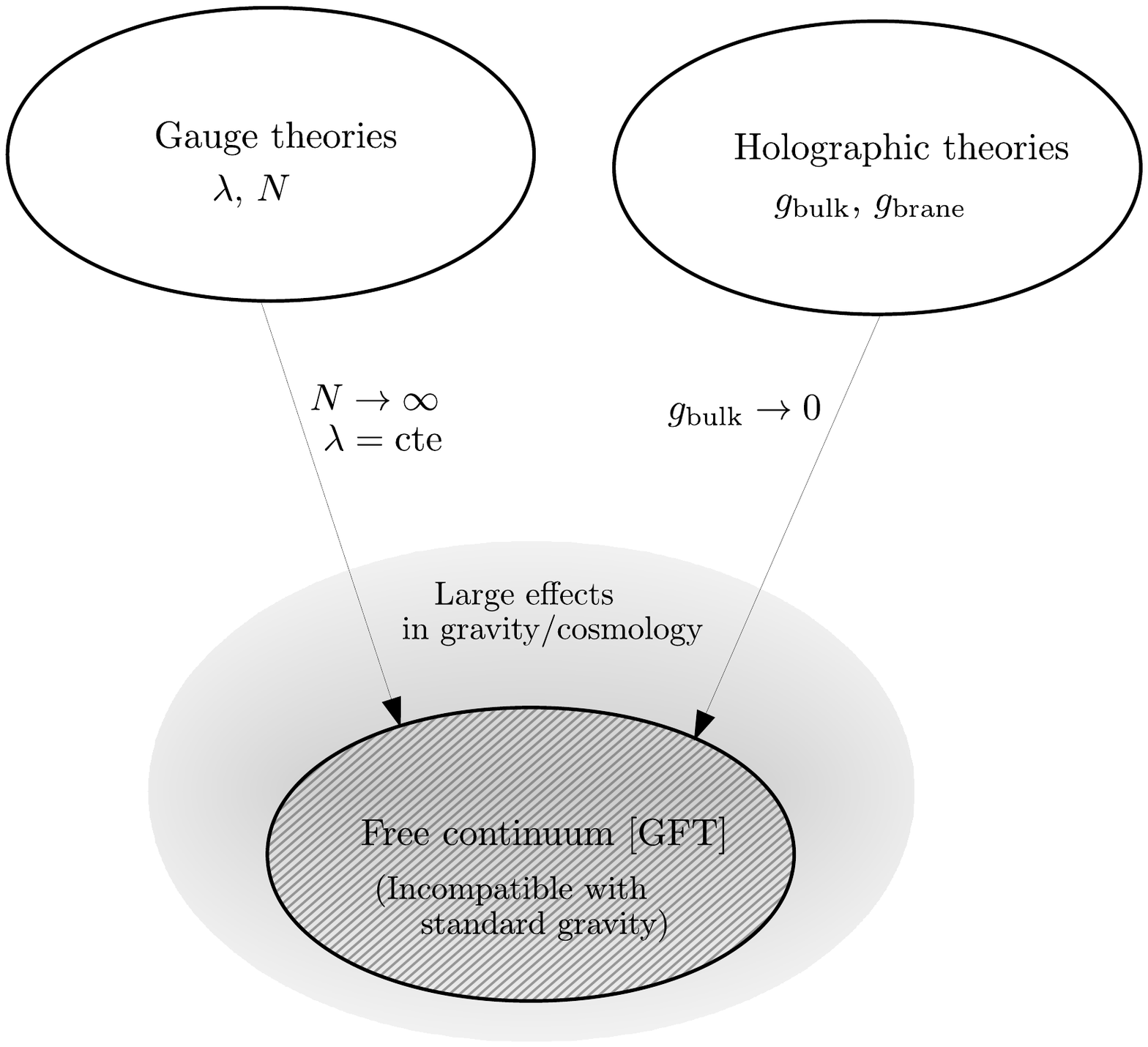}
    \caption{\it Summary of free continuum limits in the space of EFTs with gravity. See  details in sections \ref{se:models} and \ref{se:gravity}.  
    }
    \label{fig:EFT_space}
\end{figure}

\subsubsection{Continuous mass representation}
\label{se:CMR}

In the GFT the correlators of ${\cal O}$ can be described with a diagrammatic expansion using perturbation theory in the $g_{{\cal O},n}$ couplings. The resulting diagrams are built from the ${\cal O}^n$ vertices, connected by lines which are the  propagators of ${\cal O}$, \textit{i.e.}~the 2pt free correlator {$\langle{\cal O}{\cal O}\rangle_{g_{\mathcal {O},n}\to 0}$}. This is just the usual structure of Feynman diagrams, here with GFT propagators instead of ordinary propagators. 

 We can thus view the continuum sector as a set of fields  $\Phi\equiv \{\varphi_s\} $ whose only interactions are those encoded in the ${\cal O}^n$ operators. The domain for the $s$ label is determined below.  
These $\varphi_s$ fields must reproduce the propagator of ${\cal O}$.  Using the spectral representation introduced in Eq.\,\eqref{eq:spec},
 this is possible if the ${\cal O}[\varphi_s]$ operator is
\be
{\cal O}[\varphi_s] = \int^\infty_0 ds \sqrt{\rho(s)} \varphi_s \,
\label{eq:O_GFT_cont}
\ee
with
\be 
{\cal L}[\varphi_s]_{\rm continuum} \supset  \int^\infty_0 ds {\cal L}[ \varphi_s(x)]_{\rm free} \,,\quad\quad 
{\cal L}[\varphi_s(x)]_{\rm free} = -\frac{1}{2}(\partial_\mu \varphi_s)^2-\frac{s}{2}( \varphi_s)^2
\label{eq:L_GFT_cont}
\,.
\ee
The $\varphi_s(x)$ are ordinary free fields with squared mass $s$ and propagator\,\footnote{
In terms of quantization rules, one 
introduces creation and annihilation operators of fields $\varphi_s$ such that $[a_{p,s},a^{\dagger}_{p',s'}]=(2\pi)^3 2 p_0 \theta(p_0) \delta^{(3)}({\bf p}-{\bf p}')\delta(s-s')$. }
{
\be\langle\varphi_s(x)\varphi_{s'}(0)\rangle=\delta(s-s') \int \frac{d^4p}{(2\pi)^4}\frac{i e^{i p x} }{-p^2-s+i\epsilon} \,.
\label{eq:Phi_GFT_cont}
\ee
}
 Eq.\,\eqref{eq:Phi_GFT_cont} together with the definition \eqref{eq:O_GFT_cont} reproduces the spectral representation Eq.\,\eqref{eq:spec} of the  $\left\langle {\cal O}(x) {\cal O}(0) \right\rangle $ correlator. Similar developments can be found in Refs.~\cite{PhysRev.126.1209,Deshpande:2008ra}.

The higher point correlators of $\cal O$ follow trivially since they inherit the properties  of the free fields $\varphi_s(x)$. Namely, the odd correlators of $\cal O $ vanish, up to $O(g_{{\cal O},n})$, 
and the even correlators tend to the free disconnected result, up to $O(g_{{\cal O},n})$, as required for a GFT.  For example in the 4pt case, one obtains Eq.\,\eqref{eq:4pt_free}.

\section{Consistency with standard gravity \label{se:gravity} }

 In the previous section we have introduced the notion of free continuum, \textit{i.e.}~of GFT.  Here we expand the explanation of why a GFT is not compatible with 4D Einstein gravity. Throughout the paper we refer to 4D Einstein  as  \textit{standard} gravity. 
  In the present section we further often shorten ``standard gravity'' to ``gravity''.  Some of the arguments already  existed and are hereby reviewed, while others are new to the best of our knowledge.  We  then discuss gravity-compatible realizations and sketch some basic cosmological  consequences.

\subsection{Arguments from OPE }
\label{se:OPE}

In this section  we provide  arguments based on the operator product expansion (OPE).
\subsubsection{From CFT (review)}

\label{se:OPECFT}
We start with a gauge theory with  arbitrary 't\,Hooft coupling, focusing on the conformal case. In conformal theories, there is a rigorous claim  that the simultaneous existence of a generalized free field and the stress-energy tensor are incompatible, unless the generalized free field is an ordinary free field (see \textit{e.g.}~Refs.~\cite{Dymarsky:2014zja,Kap:lecture}). 

 A version of the proof of this well-known result goes as follows.
Let us assume that a conformal theory contains a generalized free field ${\cal O}$ and a stress tensor ${\cal T}_{\mu\nu}$. The stress tensor has dimension $d$ and spin $2$. The 4pt function of ${\cal O}$, given in Eq.\,\eqref{eq:4pt_free}, contains information about the spectrum and OPE coefficients of $\cal O$. It can be shown~\cite{Kap:lecture} that Eq.\,\eqref{eq:4pt_free} implies that the OPE of ${\cal O}(x){\cal O}(0)$ only contains bilinear operators built from derivatives of $\cal O$, \textit{e.g.}~${\cal O}\square^n{\cal O}$.\,\footnote{The same feature is true in an ordinary free field theory, and thus the same conclusion can be obtained using the continuum mass representation.}
Such operators have dimension {$2\Delta+2 n$}. These facts put together imply that 
 there can be a stress tensor in the OPE of ${\cal O}(x){\cal O}(0)$ only if $2\Delta+2=d$, hence requiring $\Delta=(d-2)/2$ {(\textit{e.g.}~$\Delta=1$ for $d=4$)}, which corresponds to the ordinary free field (in which case one has $\Phi(0)\Phi(x) \supset\frac{1}{2}x^\mu x^\nu {\cal T}_{\mu\nu}$). Otherwise, \textit{i.e.}~if $\Delta>1$, there cannot be ${\cal T}_{\mu\nu}$ in the OPE of ${\cal O}(x){\cal O}(0)$. The latter feature implies, by symmetry of the OPE coefficients, that ${\cal O}$ is absent of the ${\cal T}_{\mu\nu}(x){\cal O}(0)$ OPE.  This is inconsistent with translation invariance, which requires that ${\cal O}$  must appear in this OPE with a  nonzero coefficient. We thus reach a contradiction. 
 The contradiction is resolved if either the generalized free field $\cal O$ or ${\cal T}_{\mu  \nu}$ are absent from the conformal theory.

\subsubsection{From continuous mass representation}

The fact that a GFT with stress tensor is inconsistent can also be directly seen from the continuum mass representation defined in Sec.~\ref{se:CMR}. The $\rho(s)$ distribution is in general supported over $[0,\infty)$, but the argument also applies if the distribution is truncated to an interval such as $[0,\Lambda^2)$, as may occur in an EFT.  
In the presence of the free continuum described by the set of free fields $\varphi_s$, we can formally derive a stress tensor from the Lagrangian Eq.\,\eqref{eq:L_GFT_cont}, which gives ${\cal T}_{\mu\nu}=\int ds  T_{\mu\nu}[\varphi_s]$. We can then compute the correlator of this generalized free stress tensor with itself and focus on the traceless part. The result is proportional to $\int ds \,\frac{\delta(0)}{x^{2d}}$.  Since $\int ds\, \delta(0) =\infty$, the central charge is infinite.\,\footnote{This is consistent with the viewpoint of the GFT as a CFT with $N\to \infty$, which also gives an infinite central charge.} 
The infinite central charge effectively sends to zero the coefficient involving ${\cal T}_{\mu\nu}$ in the OPE of ${\cal O}(x){\cal O}(0)$. This leads to a contradiction with translation invariance, as in the CFT proof above.   
The argument given here extends beyond the CFT case, and holds whether or not there are ${\cal O}^n$ local operators.

\subsection{Arguments from species scale }
\label{se:species_scale}

Let us consider the GFT in the presence of dynamical gravity via the  action $S=S_{\rm grav}+\int d^4x \sqrt{g} \, {\cal L}_{\rm continuum}$. 
This is in general a low-energy EFT describing subPlanckian gravity interacting with matter, together with classical black holes. What is the UV cutoff scale of this EFT?  

Even though the strength of gravity is set by the reduced Planck mass $M_{\rm Pl}$, the actual validity scale of the EFT may be lower. Using an argument based on classical black hole lifetime,  Ref.\,\cite{Dvali:2008ec} established the bound $\Lambda\sim M_{\rm Pl}/\sqrt{N_{\rm sp}}$ where $N_{\rm sp}$ is the number of  species of matter in the theory. This argument relies on Hawking radiation and thus assumes that the species are stable or narrow particles. 

Here we will verify that the species bound can be extended, beyond weak coupling, to a CFT with arbitrary central charge $c$ and arbitrary `t Hooft coupling $\lambda$. This is an aside result that we use to strengthen our analysis and which is also interesting in itself. 

\subsubsection{Species scale for  CFT with arbitrary central charge }

Let us consider $S=S_{\rm grav}+\int d^4x \sqrt{g} {\cal L}_{\rm CFT}$. We want to determine the UV cutoff scale~$\Lambda$ of the theory.
Let us assume there is no cosmological constant and let us put the CFT at finite temperature $T$. The energy density is given by $\rho_{\rm CFT}=c \pi^2 \zeta  T^4$, with $\zeta=2$ and $\zeta=\frac{3}{2}$ at weak and strong coupling, respectively. For simplicity we drop the  $\pi^2\zeta $ factor in the following. 
As a result of this energy density, spacetime expands with a Hubble rate 
\be H\sim \sqrt{c}\frac{T^2}{M_{\rm Pl}}\,.\ee 
The associated volume for a Hubble patch is $1/H^3$. But this volume is bounded from below by the cutoff of the theory, as it cannot be smaller than the volume $(\Delta x)^3=1/\Lambda^3$, which amounts to a Hubble rate $H=\Lambda$. 
The corresponding momentum scale is of order $\Lambda$ and, since temperature is proportional to the  average momentum scale, we can say that this Hubble rate is attained for $T\sim \Lambda$. Therefore the UV cutoff is  determined by the condition $H|_{T=\Lambda}\sim\Lambda$, which gives
\be
\Lambda \sim \frac{M_{\rm Pl}}{\sqrt{c}}\,.
\ee
In the case of weakly coupled stable species we have $c\to N_{\rm sp}$ which recovers the usual formula from Ref.~\cite{Dvali:2008ec}.

\subsubsection{Application to GFT}

Having ensured that the species scale applies to any CFT, we turn to the GFT.
Viewing the GFT as the limit of a CFT with $c\sim N^2\to \infty$, we can see that the number of species in the GFT goes to infinity. Therefore $\Lambda\to 0$, and so there is no energy regime where gravity is weakly coupled!   This means that a GFT coupled to gravity simply does not exist. 

The same conclusion is obtained when considering the continuous mass representation. For any $\Delta>1$ there is an infinite number of degrees of freedom $N_{\rm sp}=\infty$, hence $c=\infty$, which implies $\Lambda\to 0$.

From all of the above arguments we conclude that gravity cannot couple to the GFT because the latter has infinitely too many degrees of freedom. Notice that, in contrast, a CFT has a finite number of species $c\sim N^2$, and thus in that case the UV cutoff $\Lambda$ is nonzero. 
Notice also that all the arguments would be avoided in the case of an ordinary free field ($\Delta=\frac{d-2}{2}$, $\rho(s)\propto\delta(s-m^2)$); however this is excluded in our definition of GFT (see Sec.\,\ref{se:GFT}). 
In a sense, the coupling to (4D Einstein) gravity  forces the generalized free field to be an ordinary free field. 

\subsection{Holographic theory vs GFT}

We have shown that a GFT (as defined in Sec.~\ref{se:GFT}) is obtained from a holographic setup by setting all the bulk interactions to zero. This definition implies that 5D gravity (with 5D Planck scale $M_5$) is removed when taking the GFT limit, $M_{5}\to \infty$. In such a limit we have a gravity-less bulk which can be trivially integrated out.~\footnote{A situation reproduced by Little String Theories~\cite{Aharony:1999ks}.} 
Conversely, a holographic setup with gravity automatically provides a continuum compatible with gravity.  However the price to pay is that gravity in the holographic theory is intrinsically 5D, implying that the graviton itself has a continuum component such that gravity deviates from 4D Einstein gravity.

Let us {briefly} comment about the case of AdS background (\textit{e.g.}~the RS2 setup~\cite{Randall:1999vf}). In this case the  AdS/CFT correspondence applies. From that  correspondence the $g_{\rm bulk}$ coupling goes as some power of $1/N$, and hence the GFT limit is consistent, from either the AdS or the CFT viewpoints, since $[g_{\rm bulk}\to 0] \Leftrightarrow [N\to \infty]$. We may  note that the AdS theory always has a 5D stress tensor, even without gravity. What really changes when taking $g_{\rm bulk}\to 0$ is that the graviton field is removed.  The CFT operator dual to this bulk field is the CFT stress tensor, which is thus removed upon taking $g_{\rm bulk}\to 0$. This is in agreement with the arguments of Sec.~\ref{se:OPECFT}.

\subsection{Gravity-compatible continuum models }

\label{se:gravity_compatible}

Along with the arguments of Secs.~\ref{se:OPE} and \ref{se:species_scale} we have established that a free continuum, \textit{i.e.}~a GFT, is incompatible with 4D Einstein gravity. 
We now consider theories lying in the neighborhood of this limiting case in theory space (see Fig.\,\ref{fig:EFT_space}). Such neighboring theories feature some notion of free or nearly-free continuum, and some ingredients making the continuum EFT compatible with gravity ---  associated to loopholes in the no-go arguments of Secs.~\ref{se:OPE} and \ref{se:species_scale}. 

By examining the latter arguments, we can identify the following logical possibilities for EFTs neighbors to the excluded case of GFT+4D Einstein gravity: \textit{a)} The EFT has a large, but finite, number of degrees of freedom; and, \textit{b)}  Gravity  differs from 4D Einstein gravity. 
Following these lines we then identify the  following three (possibly overlapping) classes of theories giving rise to free or nearly-free continuum models consistent with gravity.

\begin{enumerate}[label=\it\roman*)]
    \item \textit{The continuum is really a discretum.}
    
    It is possible that the free continuum be an approximation of a free discretum. Indeed both are indistinguishable to a finite precision experiment unable to resolve the discretum spacing.
    In this case the underlying degrees of freedom are countable, and their number is finite since they are bounded by a gravity-induced UV cutoff. Thus the central charge is finite and inconsistencies with gravity are avoided. In the bottom-up EFT of the free continuum this can be simply obtained by making the spectral distribution discrete in the continuous mass representation of Sec.\,\ref{se:CMR}.

\item  \textit{The continuum is a large-$N$ gauge correlator. }

  For finite number of colors $N$ the central charge is finite, thus inconsistencies with gravity are avoided. In that case the continuum is nearly-free since it has nontrivial connected correlators which are $1/N$-suppressed but nonzero. 
    At strong coupling a discretum may arise at low energy if the theory enters a confining phase, hence providing a realization of \textit{i)}.

\item \textit{The continuum is holographic.}

In this case the underlying dynamics is intrinsically 5D even though it is seen from a brane viewpoint.
The matter continuum arising in the 4D holographic theory automatically couples consistently to gravity. The counterpart is that gravity itself has a continuum component. Thus  gravity in the holographic theory is not 4D Einstein gravity.
The holographic framework can also realize the above ones in specific cases, as for certain backgrounds a discrete KK spectrum arises, hence realizing \textit{i)}, and for pure AdS background \textit{ii)} is realized via the AdS/CFT correspondence. 

\end{enumerate}

For convenience we refer to the continuum from both $i)$ and $ii)$ as a \textit{nearly-free continuum}. For $i)$, ``nearly'' applies to ``continuum'', which really is a discretum, while for $ii)$, ``nearly'' applies to ``free'', since the continuum has small but nonzero nontrivial correlators.

We can now observe that, for any kind of EFT with a free or nearly-free continuum consistently coupled to gravity, substantial deviations must appear in the gravity sector. These are the deviations that would blow up and make the theory inconsistent when taking the limit of a free continuum coupled to 4D Einstein gravity. 

This fact is evident for holographic models, class \textit{iii)}, in which gravity   automatically deviates from 4D Einstein gravity. But it also occurs in the classes of models $i)$ and $ii)$ because, in any event, the graviton propagator is dressed by insertions of $\langle T T\rangle$ correlators, which are proportional to the central charge. This is a physical QFT correction to the Newton law of gravity. In the limit of large central charge the correction to the graviton propagator blows up, inducing large effects on the gravity sector.

In summary we can state that, as a general feature, consistent  models of a free, or nearly-free, continuum must feature deviations in the gravity sector.
This is pictured in Fig.\,\ref{fig:EFT_space}.
 We investigate such effects in a concrete framework in the upcoming sections.  

\subsection{Cosmological implications}
\label{se:cosmo_implications}

In this section we qualitatively discuss the expected cosmological effects in the classes of gravity-compatible continuum models listed in Sec.\,\ref{se:gravity_compatible}. Along the same lines as the observations made there, such models must have significant impact on standard cosmology since they feature, either a large number of degrees of freedom, or deviations from gravity that blow up when approaching the forbidden limit of GFT+4D Einstein gravity (see Fig.\,\ref{fig:EFT_space}). 
Here we thus discuss basic cosmological aspects of the classes of models $i)$, $ii)$ and $iii)$, \textit{i.e.}~discretum, large $N$ gauge theories and holographic theories. 

A cosmological discretum is a fairly intuitive possibility. In that case the continuum is really made out of a set of 4D particles with standard properties, and thus their contributions to the Friedmann equation are clear. For example, at temperatures lower than the mass gap $\sigma$, the tower of particles is nonrelativistic and can be a candidate for dark matter. Such a scenario has been studied at length, see \textit{e.g.}~Refs.\,\cite{Dienes:2011ja,Dienes:2011sa}. 

The cosmological implications of a hidden large $N$ gauge theory are trickier, because in general we do \textit{not} know the equation of state $p=w\rho$, except in the following particular cases. 
First, the gauge theory may transition to a confined phase at low temperature, in which case the confined case is described by a discretum EFT already discussed above. 
Second, the gauge theory may be at a conformal fixed point, in which case it is a CFT whose properties are very constrained by symmetries. Let us review this well-known particular case. The hot CFT behaves as dark radiation because scale invariance implies $T^\mu_{\rm CFT, \mu}=0$ which in turn implies $p=\rho/3$, \textit{i.e.}~$w=1/3$. 
Since the hidden CFT has many ($\sim N^2$) degrees of freedom, the temperature $T_h$ must be much lower than the one of the visible sector, otherwise the CFT energy density $\rho_h = \zeta \pi^2 N^2 T_h^4$ overwhelms the visible one, which amounts to a too large amount of dark radiation, excluded by observations.  Hence one requires $\rho_h\lesssim \rho$. Since $N \gg 1$, such a requirement on $\rho_h$ implies that the temperature of the hidden CFT should be much lower than the visible one, 
$T_h/T_{\rm vis}\sim g^{1/4}_* N^{-1/2}\ll 1$.   
 For more general gauge theories a similar reasoning applies at a more qualitative level, yielding the  generic prediction that a large $N$ hidden sector must be \textit{ultracold} in order to not spoil cosmology. However, we cannot say more because we do not know the equation of state for such an energy density. A cosmological continuum model, apart from the CFT case, is thus best studied via holography. 

We now turn to holographic continuum models. When all Standard Model fields are identified with brane-localized modes, these are usually called ``brane-world'' in the cosmological context. The cosmology of some of these models 
has been well studied.
The simplest, and best studied, cosmological scenario is the one for which the bulk is exactly AdS everywhere, which furthermore exactly mirrors
the scenario of hot CFT reviewed above (see \textit{e.g.}\,Refs.\,\cite{Gubser:1999vj, Shiromizu:1999wj,Binetruy:1999hy,Hebecker:2001nv,Langlois:2002ke,Langlois:2003zb}).
The key point is that at finite temperature a horizon develops in the bulk, with AdS-Schwarzschild (AdSS) metric. The presence of the horizon crucially modifies the effective Friedmann equation projected on the brane with a term which, from the standpoint of the brane observer, behaves as dark radiation. This effective radiation term arising from the bulk geometry matches the CFT result $\rho_h$ for strong `t\,Hooft coupling. We summarize such a remarkable feature as 
\be
\textrm{AdS-Schwarzschild~horizon} ~~ \Leftrightarrow ~~ w_{\rm eff}=\frac{1}{3}
{~~~\textrm{(dark~radiation)} }
\,.
\ee

Departing from the pure AdS case there are plenty of possible background geometries, in particular the ``soft-wall'' backgrounds appearing from the 5D scalar-gravity system, see \textit{e.g.}~Refs.~\cite{Karch:2006pv,Gursoy:2007cb,Gursoy:2007er,Gubser:2008ny,Falkowski:2008fz, Batell:2008zm, Batell:2008me,Cabrer:2009we,vonGersdorff:2010ht, Cabrer:2011fb,Megias:2019vdb}. 
Some of these backgrounds  
give rise to a continuum in the 4D holographic theory. Continuum models from the scalar-gravity framework will be the focus of the rest of the paper.

\section{Holographic continuum: gravity and Friedmann equation}
\label{se:holcontinuum}

Our focus here  is on holographically-defined continuum models.
Such models are particularly attractive as everything is calculable since the 5D QFT is weakly coupled. 
In this section we lay out the overall framework for holographic models of continuum. The setup is reminiscent of brane-world models (see {\it e.g.}~Ref.\,\cite{Brax:2003fv}). Namely, we consider
a five-dimensional spacetime with a flat 3-brane (\textit{i.e.}~domain wall or defect) living on it, and evaluate the effective theory for an observer living in the brane worldvolume (see Fig.\,\ref{fig:background}). In such a setup the 5D excitations are integrated out and form a continuum from the standpoint of the brane observer. 

Since the overarching theme of the paper is the consistency of continuum EFT with gravity, we will be especially interested in the gravity side of the holographic continuum models. Thus two concrete objects of study stand out. 
\begin{itemize}
    \item \textit{The gravitational potential} 

At any scale for which a continuum is present in the holographic EFT, 
something nontrivial has to occur in the gravity sector to ensure consistency with gravity. Thus some deviation from Newtonian gravity can be expected at such scales.  This can also be qualitatively understood in terms of the existence of a stress tensor in the continuum sector.  Such a stress tensor, whose existence is ensured in the holographic setup, 
dresses the 4D graviton,  yielding a modification of the Newtonian potential.

    \item \textit{The Friedmann  equation}
  
    The equation of state in the continuum sector is in general nontrivial and unknown (see also the discussion in Sec.\,\ref{se:cosmo_implications}). However, in holographic models this equation of state is encoded into the \textit{geometry} of the 5D background. This appears at the level of the effective 4D Friedmann equation seen by a brane observer, which contains nontrivial information about the bulk geometry --- and thus about the equation of state. 
    The 4D holographic theory should also satisfy consistency conditions imposed by the  5D continuity equation projected onto the brane.
    
\end{itemize}
In summary we expect deviations to both the Newtonian potential and the Friedmann equation.

\subsection{The five-dimensional background }

We consider a five-dimensional spacetime with a flat 3-brane (\textit{i.e.}~domain wall). 
The 5D coordinates are denoted by upper case roman indices $M,N,\ldots$, while the 4D coordinates on the 3-brane $\mathcal M$ are denoted by greek $\mu,\nu,\ldots$ indices. 

We consider the action of the scalar-gravity system 
\be
S= \int d^5x \sqrt{g} \left(\frac{M^3_{5}}{2}\, {}^{(5)}R -\frac{1}{2}(\partial_M \phi)^2 -V(\phi)\right) -
\int_{\rm brane}  d^4x \sqrt{\bar g} \, \left(V_b(\phi)+\Lambda_b\right)~+S_{\rm matter}+\ldots
\,, \label{eq:S_DG}
\ee
with $\phi$ the scalar (dilaton) field,  {$^{(5)}R$ the 5D Ricci scalar}, $M_5$ the fundamental 5D Planck scale, $\Lambda_b$ the brane tension, $\bar g_{\mu\nu}$ the induced metric on the brane,  $V$ and $V_b$ the bulk and brane-localized potentials for $\phi$. $S_{\rm matter}$ encodes the action for quantum fields living on this background.  
The ellipses encode the Gibbons-Hawking-York term~\cite{York:1972sj,Gibbons:1976ue}.
  We assume that the brane potential sets the scalar field vacuum expectation value (vev) to a nonzero value $\langle\phi\rangle=v_{\brane}$, with $V_{\brane}(v_{\brane})=0$. The bulk potential is model dependent and explicitly given further below.

The general ansatz for the  5D metric is 
\begin{align}
ds^2 = g_{MN} x^Mx^N & = \omega^2(z)\left( - f(z) d\tau^2 + d \x^2 +\frac{1}{f(z)}dz^2 \right) \label{eq:ds2conf}  \\
& = -n^2(r)d\tau^2 + \frac{r^2}{\ell^2} d\x^2 + b^2(r) dr^2 \label{eq:ds2brane}
\,.
\end{align}
The coordinate frame in the first line shows that this metric is conformally related to the flat space Schwarzschild metric. The functions $\omega(z)$ and $f(z)$ are referred to, respectively, as the warp and blackening factors. 
The coordinate frame in the second line is convenient for brane cosmology. 
Along any constant slice of $r$, the $r=\ell \omega(z)$ coordinate acts like a {cosmological} scale factor.

The 3-brane is a hypersurface located at $r=r_{\brane}$. 
With the above coordinates the induced metric $\bar g_{\mu\nu}$ reads as
\be
d\bar s^2=\bar g_{\mu\nu}dx^\mu dx^\nu=-d t^2 + \frac{r_{\brane}^2}{\ell^2}d\x^2\,,
\ee
where we have introduced the brane cosmic time $dt=n(r_{\brane})d\tau$. 
According to this metric, if the brane moves along $r$ in the extra dimension, \textit{i.e.}~if $r_{\brane}=r_{\brane}(t)$, the brane observer perceives expansion of the 4D universe with Hubble factor $H=\dot r_b/r_b$, where $\dot r_b=\partial_t r_b$.

The 5D equations of motion for metric factors and the dilaton field are given in the cosmological frame by~\cite{Megias:2018sxv} 
\begin{eqnarray}
\hspace{-0.3cm}&& \frac{n^{\prime\prime}(r)}{n(r)} - \left(\frac{n^\prime(r)}{n(r)} - \frac{1}{r}  \right) \left( \frac{b^\prime(r)}{b(r)} - \frac{2}{r} \right) = 0 \,,  \label{eq:EoM1}  \\
\hspace{-0.3cm}&&\frac{n^\prime(r)}{n(r)} + \frac{b^\prime(r)}{b(r)} - r \bar\phi^\prime(r)^2 = 0 \,,   \label{eq:EoM2}
\\
\hspace{-0.3cm}&&\frac{n^\prime(r)}{n(r)} + \frac{1}{r} +  r\, b^2(r) \bar V(\bar\phi) - \frac{r}{2}\bar\phi^\prime(r)^2 = 0 \,,   \label{eq:EoM3} \\
\hspace{-0.3cm}&& \bar\phi^{\prime\prime}(r) + \left( \frac{n^\prime(r)}{n(r)} - \frac{b^\prime(r)}{b(r)} + \frac{3}{r} \right) \bar\phi^\prime(r) -  b^2(r) \frac{\partial \bar V}{\partial \bar\phi} = 0 \,, \label{eq:EoM4}
\end{eqnarray}
where for convenience we have defined the dimensionless scalar field~$\bar\phi \equiv \phi / (\sqrt{3} M_5^{3/2})$, and the reduced potential $\bar V \equiv V / (3 M_5^3)$. The general solutions contain five integration constants. However it turns out that one of the equations, \textit{e.g.}~Eq.\,\eqref{eq:EoM3}, acts as an \textit{algebraic} constraint on the integration constants, hence there are only four independent constants. Some of the integration constants have no physical meaning and can be fixed without loss of generality \textit{i.e.}~amount to ``gauge redundancies'', while  others have a physical meaning. A detailed discussion is provided {for the different models} in App.~\ref{app:Technical_details}.

\subsection{The effective Friedmann equation } 

\label{se:Friedmann}

The effective Einstein equation seen by an observer standing on the brane is computed from the 5D Einstein equation, projected on the three-brane via the Gauss equation together with the Israel junction condition, which relates the extrinsic curvature to the brane-localized stress tensor. To perform the projection one  introduces the unit vector $n_M$ normal to the brane and outward-pointing,  that satisfies  $n_M n^M = 1$ and  $\bar g_{MN} = g_{MN} - n_M n_N$. 
The effective Einstein equation takes the form\,\cite{Shiromizu:1999wj, Tanaka:2003eg}
\begin{align}
R_{\mu\nu}-\frac{1}{2}g_{\mu\nu}R = \frac{1}{M^2_{\rm Pl}} \left( T^b_{\mu\nu} + T^{\rm eff}_{\mu\nu} \right) +  \frac{1}{M^6_5} \pi_{\mu\nu}  
\,. \label{eq:brane_EE}
\end{align}
Here 
$R_{\mu\nu}$ is the Ricci tensor projected on the brane. $T^b_{\mu\nu}$ is the stress energy tensor of brane-localized matter.
The $\pi_{\mu\nu}$ tensor is a quadratic combination of brane-localized stress tensors, thus
it comes with a $M^{-6}_5$ factor. 
Finally, $T^{\rm eff}_{\mu\nu} $ is the ``holographic'' effective stress tensor encoding nontrivial effects from the bulk,
\be T^{\rm eff}_{\mu\nu} = \tau^{W}_{\mu\nu}+\tau^{\phi}_{\mu\nu}+\tau^{\Lambda}_{\mu\nu}\,.
\label{eq:Teff}
\ee
The three terms are:
\\ 
\textit{i)} The projection of the 5D Weyl tensor  ${}^{(5)}C^{M}{}_{NPQ}$  on the brane,
\begin{align}
\frac{1}{M^2_{\rm Pl}} \tau^{W}_{\mu\nu} = - {}^{(5)}C^{M}{}_{NPQ} n_M n^P \bar g_\mu{}^{N} \bar g_\nu{}^Q \,
\end{align}
\textit{ii)} The  projection of the bulk stress tensor,
\begin{align}
\frac{1}{M^2_{\rm Pl}} \tau^{\phi}_{\mu\nu}=\frac{2}{3 M_5^3} \left[ T^\phi_{MN} \bar g_{\mu}{}^{M} \bar g_{\nu}{}^{N} + \big(T^\phi_{MN} n^M n^N - \frac{1}{4} T^{\phi,M}_{M} \big) \bar g_{\mu\nu} \right] \,
\end{align}
\textit{iii)} The contribution from the brane tension,
\begin{align}
\frac{1}{M^2_{\rm Pl}} \tau^{\Lambda}_{\mu\nu}=-\frac{\Lambda_b^2}{12M_5^6} \bar g_{\mu\nu}\,.
\end{align}
The brane tension is ultimately tuned in order to set the  effective 4D cosmological constant to zero.

The $\pi_{\mu\nu}$ tensor appearing in Eq.\,\eqref{eq:brane_EE} is a term induced by the extrinsic curvature terms in the Gauss equation and contains quadratic combinations of the brane stress tensor $(T^b_{\mu\nu})^2$. We can see that, if the components of the brane stress tensor satisfy
\be 
|T^b_{\mu\nu}|\ll\frac{M^6_5}{M^2_{\rm Pl}}\label{eq:smallT}\,,
\ee
the effect of $\pi_{\mu\nu}$ is negligible with respect to the standard $M^{-2}_{\rm Pl} T_{\mu\nu}$ term of Einstein equation.  In this low-energy regime, all of the physics from  the bulk    is encoded into the effective stress tensor $T^{\rm eff}_{\mu\nu}$. We always use this low-energy restriction throughout our computation. It is our  only simplifying assumption.

We plug  the metric of Eq.\,\eqref{eq:ds2brane} into Eq.\,\eqref{eq:brane_EE}, for a brane  at $r=r_{\brane}$.
We tune the 4D cosmological constant, focusing on the $(0,0)$ component of Eq.\,\eqref{eq:brane_EE} and using $\tau_{00}=\rho$.
Assuming $\rho\ll M^6_5/M^2_{\rm Pl}$ we obtain the effective Friedmann equation on the brane
\be
3 M^2_{\rm Pl} \left(\frac{\dot r_{\brane}}{r_{\brane}}\right)^2 = \rho_{\brane} + \rho_{\rm eff}(r_b) + O\left(\frac{\rho_{\brane}^2 M^2_{\rm Pl}}{M^6_5} \right)\,, \label{eq:EFE}
\ee
where $\rho_{b}$ is the energy of brane-localized matter and
\begin{equation}
\rho_{\rm eff} =\rho^W(r_b)+\rho^\phi(r_b)+ \rho^\Lambda  \label{eq:rhoeff}
\end{equation}
encodes the contributions from Weyl tensor, bulk scalar stress tensor and brane tension, following Eq.\,\eqref{eq:Teff}. As indicated in Eq.\,\eqref{eq:rhoeff},  the bulk contributions depend in general on the brane location $r_b$. 
Since the dependence in $r_b$ amounts to the dependence in the  cosmological scale factor, it  determines the ``equation of state'' of the various terms in $\rho_{\rm eff}$. 
In section \ref{se:holcontinuum} we compute $\rho_{\rm eff}$ for specific backgrounds.

\subsubsection{The Weyl energy}

While the $\tau^\phi$ term is a fairly intuitive contribution from the bulk stress tensor, the $\tau^W$ term  comes purely from  bulk geometry. We have thus a geometric effect from the bulk that induces  an effective energy term on the brane.   We refer to $\rho^W$ as the Weyl energy. 
The Weyl energy   depends on the $C_{5050}$ component of the Weyl tensor, 
\begin{eqnarray}
\rho^W(r_b) &=&- \frac{M^2_{\rm Pl}}{n^2(r_{\brane})b^2(r_{\brane})}C_{5050}(r_{\brane})  \nonumber \\
              &=&- \frac{M^2_{\rm Pl}}{2 b(r_{\brane})^2}  \left[ \frac{n^{\prime\prime}(r_{\brane})}{n(r_{\brane})} - \left( \frac{b^\prime(r_{\brane})}{b(r_{\brane})} + \frac{1}{r_{\brane}} \right) \left( \frac{n^\prime(r_{\brane})}{n(r_{\brane})} - \frac{1}{r_{\brane}} \right)   \right] \,. \label{eq:Cr0_general}
\end{eqnarray}
The second line is the result obtained in the brane cosmology coordinates of Eq.~(\ref{eq:ds2brane}). 

In general, the Weyl tensor is a measure of deviation from conformality.  It vanishes if $f\to 0$ identically in the conformal coordinates of Eq.\,\eqref{eq:ds2conf}. The Weyl tensor is instead nonzero in the presence of a horizon. In the conformal frame, if $f$ vanishes at a given point $f(z_h)=0$, the hypersurface $z=z_h$  is a horizon whose temperature  is given by $T_h = |f^\prime(z_h)|/(4\pi)$.~\footnote{The expression of the temperature in the brane cosmology coordinates of Eq.~(\ref{eq:ds2brane}) is $T_h = \frac{1}{4\pi} \sqrt{\frac{d n^2(r)}{dr} \frac{d \chi^2(r)}{dr}} \Big|_{r = r_h}$ where $\chi(r) \equiv 1/b(r)$.} The presence of the Weyl energy in the brane Friedmann equation is thus associated with the temperature of the horizon in the bulk. In Sec.\,\ref{se:lineardilaton} we  compute the horizon temperature for completeness and to compare with the literature.\,\footnote{
The entropy  of the horizon is given by $\mathcal S =V_3\, \omega^3(z_h)/(4 G_5)$, where  $V_3$ is the volume in  3D space and $G_5$ is the 5D Newton constant. In the brane cosmology coordinates we find  $\mathcal S = V_3 \left(  \frac{r_h}{\ell} \right)^3 / (4 G_5)$  for any model.}
 
 \subsubsection{The effective conservation law}

Even though  the 5D physics is projected onto the brane, the 5D conservation law leaves a nontrivial imprint on the resulting 4D physics. Combined  with the 4D Bianchi identity it produces a nontrivial conservation law for our effective energy term \cite{Langlois:2003zb,Tanaka:2003eg},
\begin{equation}
\dot \rho_{\rm eff} + 4 H \rho_{\rm eff} + H T^{{\rm eff}\,\mu}_\mu = - 2 \left( 1 + \frac{\rho_{\brane}}{\Lambda_{\brane}}\right) T^\phi_{MN} u^M n^N   \,. \label{eq:5Dcont}
\end{equation}
Here $u^M$ is the timelike unit  vector for brane observers, $u^M= (\dot \tau,{\bm 0}, \dot r_b)$ in the cosmological frame. 
Notice the $4H$ factor arising due to 5D spacetime. 
To obtain this form we have used that 
\begin{equation}
2  \frac{M^2_{\rm Pl}}{M_5^3}H T^\phi_{MN} n^M n^M + H\tau_\mu^{\Lambda\,\mu} = H T^{{\rm eff}\,\mu}_\mu\,.
\label{eq:modified-japanese}
\end{equation}
Note that since where $\tau_{\,\, \mu}^{{\rm \Lambda}\, \mu} = -4 \rho^\Lambda$, the $\rho^\Lambda$ terms cancel inside this conservation equation.

  The energy density of brane  matter $\rho_{\brane}$ and the
brane tension  $\Lambda_{\brane}$ have been defined earlier in this section. The $\rho_b/\Lambda_b$ term is negligible in the low energy regime defined by Eq.\,\eqref{eq:smallT}. 
For our purposes, the conservation equation \eqref{eq:5Dcont} serves as a highly nontrivial consistency check of the cosmological framework.

\subsection{QFT overview }

\label{se:QFT}

In this section we consider quantum fields living over the 5D background encoded in the term $S_{\rm matter}$ in Eq.~(\ref{eq:S_DG}).   
We review some essential properties of bulk QFT, as seen from a brane, that are needed to establish the general picture  of a holographic continuum model. 

Our focus is on the fields living in the $r\leq r_{\brane}$, \textit{i.e.}~$z\geq z_{\brane}$, region of the bulk.  
We assume that the fields have Neumann boundary conditions (BC) on the brane, \textit{i.e.}~the fields are allowed to fluctuate on the brane.\,\footnote{
A field with Dirichlet boundary condition would contribute to the brane correlators only via internal lines. This is not the focus of the present study. 
} 
The fields are described by a Lagrangian in the 5D bulk, but additionally there can always be operators localized on the brane. In fact those are always generated by loop effects (see Ref.\,\cite{Fichet:2021xfn} for explicit results). Thus following the EFT paradigm such operators should be included in the brane Lagrangian in a first place.

Let us now consider a generic bulk field $\Phi$ with value $\Phi_0\equiv \Phi|_{\rm brane}=\Phi(r_b)$ on the brane. The field propagates in the bulk, but the brane-localized operators would influence its propagation. In fact, on  general grounds, a brane-to-brane propagator 
takes the form $G=[G^{-1}_0+{\cal B}]^{-1}$, where $G_0\equiv G|_{{\cal B}=0}$ and ${\cal B}$ is the  bilinear insertion induced by the brane-localized operators~\cite{Fichet:2021xfn}, and dressing $G_0$.  
In momentum space, both ${\cal B}$ and possibly $G^{-1}_0$ contain an analytic piece $\propto p^2$, which amounts to having an isolated 4D free mode in the spectrum. This mode is tied to the fluctuation of the field on the brane, and its wavefunction is typically localized near the brane. Singling this 4D localized mode out, the propagator can be written as 
\begin{align}
\langle
\Phi_0(x^\mu)\Phi_0(0) \rangle  = \int \frac{d^4p}{(2\pi)^4} e^{i p\cdot x} G(-p^2) \,,\quad\quad
G(q^2)=\frac{i Z_0}{ q^2 -m_0^2 + b_0 \Pi(q^2) }\,, \label{eq:b2b}
\end{align}
where $Z_0$ is a wavefunction renormalization effect, and the $\Pi(q^2)$ term is non-analytical. For sufficiently smooth background, as the one we will consider here, $\Pi(q^2)$ has a branch cut along some region of the $q^2 > 0$ axis,  \textit{i.e.}~it is a continuum.   This term encodes the contribution of all the rest of the bulk modes to the brane-to-brane propagator.  We thus have split the denominator into a 4D free piece and a continuum piece. 

We can see that the structure of Eq.\,\eqref{eq:b2b} amounts to the one of a 4D free propagator dressed by insertions due to mixing with a continuum (see Fig.\,\ref{fig:background}).  This is the same structure as the  $\langle\varphi(x)\varphi(0)\rangle$ propagator of the continuum EFT in Eq.\,\eqref{eq:L_eff} dressed by $\langle {\cal O} {\cal O}\rangle$ insertions, upon identifying ${\tilde O}[\varphi] \propto \varphi$ and $\langle{\cal O}{\cal O}\rangle\propto \Pi$. 
\footnote{The notion of mixing  can be understood more explicitly as follows.  In the set of all degrees of freedom of $\Phi$, we can single out those which do not fluctuate on the brane, \textit{i.e.}~have Dirichlet BC. Writing 
$\Phi=\Phi_0 K_{q}(z)+\int_\lambda\Phi^\lambda_D f^\lambda(z)$, with $K_{{q}}\equiv K(q^2)$ the amputated brane-to-bulk propagator and $f_D^\lambda$ the continuous basis of Dirichlet modes, 
the set $(K_{q},f^\lambda_D)$ forms a complete basis which is orthogonal --- in the sense that the quadratic action is diagonal in $(\Phi_0,\Phi^\lambda_D)$~\cite{Fichet:2021xfn}. In this basis $\Phi_0$ has a nontrivial propagator, Eq.\,\eqref{eq:b2b}, \textit{i.e.}~a nontrivial spectral distribution. 
However one could instead, as introduced in Ref.~\cite{Batell:2007jv}, trade the 
$K_{q}(z)$ component for $K_{q=m_0}(z)$, in which case the associated degree of freedom $\Phi_0\equiv\varphi$ simply is a 4D free field. 
In that case the propagator of $\varphi$ is trivial, but in counterpart the $(\varphi,\Phi_D)$ basis is not orthogonal (see \cite{Batell:2007jv}),  and therefore there is a mixing between $\varphi$ and $\Phi_D$.  The form of Eq.\,\eqref{eq:b2b} is understood as a manifestation of this mixing. 
}

In summary, we obtain that the holographic setup leads to a continuum model that is described by the generic continuum EFT of section \ref{se:models}.
The crucial gain with respect to the generic continuum Lagrangian Eq.\,\eqref{eq:L_eff}  is that, here, the setup dictates exactly  how the law of gravity is modified.  Before focusing on gravity we discuss qualitatively some other QFT aspects which are useful for the overall understanding of the model.

\subsubsection{Spectrum and continuum final state}

The continuum piece $\Pi(q^2)$ may, or may not, be supported at the pole location given by $q^2-m^2_0+b_0\Pi(q^2)\equiv 0$. In analogy with familiar weakly coupled QFT we can distinguish two cases. If the pole lies in a region where $\Pi(q^2)$ is zero, the 4D mode described by the propagator Eq.\,\eqref{eq:b2b} is stable. It thus contributes as a Dirac delta function to the spectral distribution and is identified as a particle in the Hilbert space of the 4D theory. In contrast, if the pole lies in a region where 
$\Pi(q^2)$ is nonzero, the 4D mode acquires a width $\Gamma$ given by $m_0\Gamma={\rm Im}\, \Pi(m_0^2)$ and thus amounts to a resonance, as first noted  in Ref.~\cite{Dubovsky:2000am}. This striking feature means that the 4D mode has a nonzero probability to convert into the continuum.

We notice here a key difference between continuum and discretum. If $\Pi$ was a discretum, \textit{e.g.}~$\Pi(q^2)\sim \sum_i a_i \delta(q^2-m_i^2)$, the isolated 4D mode would remain exactly stable. Such a propagator would simply describe a mixing between the 4D mode and the discretum.  
This, in a sense, is because a free particle cannot just convert into another one with different mass. 
In contrast, the mass of the continuum is a continuous variable, thus it can be arbitrarily close to $m_0$. As a result  there is a well-defined probability for the 4D mode to convert into the continuum.\,\footnote{
At a deeper level, a continuum does not have the properties required to build the familiar asymptotic multiparticle states of flat space, and may thus obey other rules. In the AdS case, for example, the continuum amounts to the normalizable bulk modes of AdS, that we know are perfectly stable (see  \textit{e.g.}\,Ref.\,\cite{DiPietro:2021sjt}).
Diagrams with AdS modes, such as $1\to 2$, for example, only induce a mixing of the bulk modes, and thus amount in familiar terms to a radiation process rather than a decay process that would remove the initial mode from the spectrum (see \textit{e.g.}~\cite{Fichet:2021pbn}).}

\begin{figure}[t]
    \centering
    \includegraphics[width=0.8\linewidth,trim={4cm 2cm 0cm 2cm},clip]{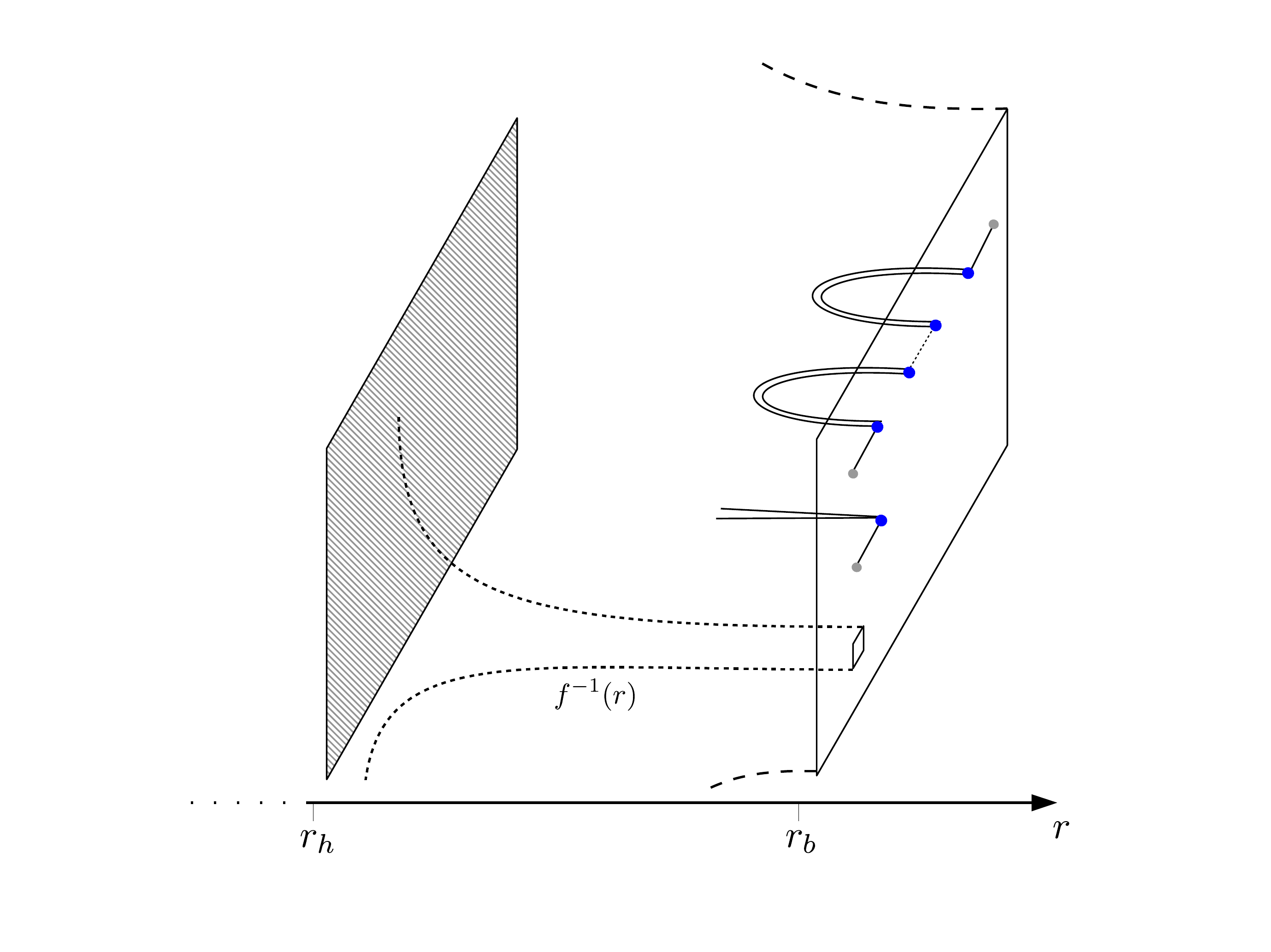}
    \caption{\it Overview of the holographic continuum framework in the  cosmological frame.  The brane and the  horizon are located respectively at $r=r_b$ and $r=r_h$. The inverse blackening factor  is pictured by the dotted line. The brane effective action features  isolated 4D modes, identified in the analytical part of the self-energy. These are represented by the simple lines on the brane. The non-analytical part of the self-energy corresponds to the bulk modes, here represented by the double lines.  
  The mixing between the modes is represented by the blue vertices. In particular the 4D graviton mode is dressed by a continuum component, which in turn implies a deviation from the Newtonian potential. The production of bulk modes from the brane feeds the horizon in the bulk, which in turn holographically induces an effective energy term in the brane Friedmann equation.  
    }
    \label{fig:background}
\end{figure}

The spectral function contains the necessary information to describe a continuum final state. In practice, in a given diagram one can simply take a unitarity cut on the generic brane-to-brane propagator, Eq.\,\eqref{eq:b2b}.
In particular, in the case of a stable particle the result takes the form 
\be
{\rm Disc}_{s}[G(s)]=Z_0 \left( 2\pi\delta(s-m^2_0) - i \frac{ b_0 {\rm Disc}[\Pi(s)]}{|s -m_0^2 + b_0 \Pi(s)|^2} \right)\,, \label{eq:DiscGp}
\ee
where ${\rm Disc}_s$ computes the discontinuity across $s>0$, as defined in Sec.~\ref{se:gravpot}. In Eq.\,\eqref{eq:DiscGp}, 
${\rm Disc}_{s}[G(s)]$ is real and ${\rm Disc}_{s}[\Pi(s)]$ is imaginary.  
We can see from Eq.\,\eqref{eq:DiscGp} that the final state can, either be the stable 4D mode, or transition via a 4D propagator into the continuum. 
In the notation of the generic Lagrangian of Eq.\,\eqref{eq:L_eff}, this amounts to a ``$\varphi^*\to \;$continuum" process, see Fig.\,\ref{fig:background}.

\subsubsection{Finite temperature}

The sector of brane-localized 4D modes can form a thermal bath. In such a case we can simply say that there is finite temperature on the brane. 
The conversion processes highlighted in the above section appear in the collision term of the Boltzmann equation of the 4D modes. They describe a sustained flux of radiation into the continuum of bulk modes, dumping energy into the bulk. In a sense these processes are responsible for ``heating up'' the bulk since, when falling deep enough in the bulk, they create a horizon which is encoded in the blackening factor $f(z)$ of Eq.\,\eqref{eq:ds2conf}  (see \textit{e.g.}~Ref.\,\cite{Hebecker:2001nv}).  
Such processes, and the overall coupled dynamics, have been studied in a number of references at various degrees of refinement using both the 5D and dual 4D viewpoints, see \textit{e.g.}~Refs.\,\cite{Gubser:2002zh, Hebecker:2001nv,Langlois:2002ke,Langlois:2003zb,Brax:2019koq,Redi:2020ffc}. 
Similar calculations could be done in the linear dilaton background, although this is not the {main} focus of the present work. 

In the present work we use the fact that  the  brane-to-bulk conversion processes quickly lose efficiency when the temperature drops (see~\cite{Gubser:2002zh, Hebecker:2001nv,Langlois:2002ke,Langlois:2003zb,Brax:2019koq,Redi:2020ffc}). We  focus on a low-energy regime in which the initial $r_h$ is a free parameter --- which encapsulates the effect of all the energy previously dumped into the bulk. We will see in Sec.~\ref{se:lineardilaton} that, in the considered holographic continuum models, $r_h$ is either  constant 
or evolves as a function of the brane location $r_b$ along the cosmological history.

\subsection{Gravitational potential}

\label{se:gravpot}

The  graviton propagator should, following the above discussions, describe a massless 4D mode with bilinear mixing to a continuum.
We denote the general propagator as
\begin{align}
\langle h_0^{\alpha\beta}(p)h_0^{\rho\sigma}(-p)\rangle =  G^{\bf 2}_{-p^2} \, \theta^{\alpha\beta\rho\sigma} \, \label{eq:b2bgrav}
\end{align}
where $G^{\bf 2}_{-p^2}=G^{\bf 2}(-p^2)$ and the superindex ${\bm 2}$  refers to the spin of the graviton.
The polarization structure $\theta^{\alpha\beta\rho\sigma}$ is given  below. 
  What is the gravitational potential resulting from the propagator, Eq.\,\eqref{eq:b2bgrav}? 
To obtain it we write the spectral representation of the propagator as~\cite{Zwicky:2016lka}
\begin{align}
\langle h_0^{\alpha\beta}(p)h_0^{\rho\sigma}(-p)\rangle
  &=
  \frac{1}{2\pi i}
  \int_0^\infty 
  ds\, 
  \frac{ 
    \text{Disc}_{s}\left[G_{s}^{\bf 2}\right] 
  }{
    s+p^2 - i \epsilon
  }
\theta^{\alpha\beta\rho\sigma}_{s} \,,
\label{eq:spectral:discontinuity}  
\end{align}
where $\text{Disc}_{s} \left[g(s)\right]$ is the discontinuity of $g(s)$ across the 
branch cut along the real line, $s \in \mathbbm{R}^+$:
\begin{align}
  \text{Disc}_{s}[g(s)]
  &=\lim_{\epsilon\to 0}\, \left( g(s+i\epsilon)-g(s-i\epsilon) \right) \,,
  & \epsilon
  &>0
  \ .  
\end{align}
In this representation the tensor structures are 
those of the standard Fierz-Pauli propagator~\cite{Hinterbichler:2010es}, 
\begin{align}
& \theta^{\alpha\beta\rho\sigma}_{s=0} = \frac{1}{2}\left(
\eta^{\alpha\rho} \eta^{\beta\sigma}+
\eta^{\alpha\sigma} \eta^{\beta\rho}-
\eta^{\alpha\beta} \eta^{\rho\sigma}
\right) \,, \\
& \theta^{\alpha\beta\rho\sigma}_{s>0} = 
\frac{1}{2}\left(
P^{\alpha\rho} P^{\beta\sigma}+
P^{\alpha\sigma} P^{\beta\rho} \right)
-\frac{1}{3}
P^{\alpha\beta} P^{\rho\sigma} \,,
\end{align}
with $P^{\alpha\beta}=\eta^{\alpha\beta}-\frac{p^\alpha p^\beta}{s}$. 

The potential can be directly obtained using the spectral representation Eq.\,\eqref{eq:spectral:discontinuity} (see \textit{e.g.}~\cite{Callin:2004py} and also\,\cite{Chaffey:2021tmj}). We pick point sources at rest such that $T_{1,2}^{\alpha\beta}=m_{1,2}\delta^{\alpha}_0\delta^{\beta}_0$. Performing the $d^3\mathbf{q}$ integral yields a general representation of the long-range potential as
\begin{align}
  V_N(R)
  &= 
  - \frac{m_1m_2}{\pi M^2_{\rm Pl}  }
  \int_0^\infty ds \;
  \text{Disc}_{s}\left[G_{\sqrt{s}}\right] 
  \;
  \frac{e^{-\sqrt{s}R}}{R} \theta^{0000}_{s}
  \,, 
\label{eq:V_spectral_gen}  
\end{align}
where $\theta^{0000}_{s=0}=\frac{1}{2}$, $\theta^{0000}_{s>0}=\frac{2}{3}$. 

If $G_{-p^2}^{\bf 2}=\frac{i}{-p^2+i\epsilon}$ we have  $\text{Disc}_{s}[\frac{i}{s+i\epsilon}]=2\pi \delta(s)$, which reproduces the standard Newtonian potential. The graviton propagator of our focus features a continuum term, structurally analogous to Eq.\,\eqref{eq:b2b}.  This term induces  a deviation from the Newtonian  potential. In Sec.~\ref{se:lineardilaton} we will compute explicitly this continuum-induced deviation in specific 5D backgrounds.

\section{The holographic gapped continuum}
\label{se:lineardilaton}

The scalar-gravity background is fixed by the choice of the bulk potential $V(\phi)$ (or, equivalently, by an associated superpotential, see \textit{e.g.}~\cite{Cabrer:2009we,Megias:2019vdb} ).
Here we mainly focus on a specific version of the scalar-gravity setup 
called the linear dilaton (LD) and a deformed version of it.
A general solving of the scalar-gravity system in these cases is presented in App.\,\ref{app:Technical_details}. 

The linear dilaton background has the fascinating property that it naturally realizes the notion of a \textit{gapped continuum} that was proposed phenomenologically in Ref.\,\cite{Georgi:2007ek}.  
Throughout this work  we assume the presence of a thermal bath on the brane, which induces a horizon in the bulk via QFT processes as described in Sec.\,\ref{se:QFT} and Fig.\,\ref{fig:background}.  

\subsection{AdS-Schwarzschild (review)}

\label{se:AdSS}

As a warm-up we briefly revisit the  AdS-Schwarzschild background. 
 The well-known case of pure AdS background is recovered in the case where {$V(\phi) = - 6 M_5^3/\ell^2$} with $M_5^3 = M_{\rm Pl}^2/\ell$, and the dilaton has no VEV, \textit{i.e.}~$\phi(z) =$ const. For $f\neq 1$ the background is AdS-Schwarzschild, \textit{i.e.}~hot AdS.  In the cosmological context this amounts to the RS2 model~\cite{Randall:1999vf} at finite temperature. In that case one has, in conformal coordinates
\be
f_{\AdSS}(z) =1-\frac{z^4}{z_h^4} \,, \quad \omega_{\AdSS}(z)=\frac{\ell}{z} \,,
\ee
 for any value of $z$, where $z_h$ is the location of the horizon, and, in brane cosmology coordinates
 \be
n_{\AdSS}(r)=\frac{r}{\ell} \sqrt{ 1-\frac{r^4_h}{r^4} } \,,\quad b_{\AdSS}(r)=\frac{\ell}{r} \frac{1}{ \sqrt{1-\frac{r^4_h}{r^4} } }\,,
 \ee
where $r_h$ is the corresponding location of the horizon, and we have used the relation $\frac{\ell}{z}=\frac{r}{\ell}$. Finally, the temperature of the black hole is
\begin{equation}
T_{\AdSS} = \frac{r_h}{\pi \ell^2} \,.
\end{equation}

\subsubsection{Deviation from the Friedmann equation}

We  find the components of the effective energy density
\be
\rho^{W}_{\AdSS}(r_{\brane})=\frac{3M^2_{\rm Pl}}{\ell^2}\frac{r^4_h}{r_{\brane}^4}\,,\quad\quad\quad\quad  \rho^{\phi}_{\AdSS} + \rho^{\Lambda}_{\AdSS} = 0 \,.
\ee
In this case $\rho^{\phi}_{\AdSS}$ is constant and cancelled by tuning the brane tension to $\Lambda_b=6\frac{M^2_{\rm Pl}}{\ell^2}$, setting the 4D cosmological constant to zero. The Weyl energy is regular at the Schwarzschild horizon.  
The total effective energy density is simply $\rho^{\rm eff}_{\AdSS}=\rho^{W}_{\AdSS}$.
The scaling in $r_b$ with the universe scale factor, $r_b=\ell a$ indicates that  the effective energy behaves as 4D radiation. This is the standard result, in accordance with the discussion in Sec.~\ref{se:cosmo_implications}.

\subsubsection{Deviation from the Newtonian potential}

In AdS the reduced brane-to-brane graviton propagator takes the  form (see \textit{e.g.}\,Ref.\,\cite{Fichet:2019owx})
\be
G^{\bm 2}_{{\rm AdSS}}(-p^2) =
 -\frac{i}{p^2}+i\left(2\gamma-1+2\log\left(\sqrt{p^2}\frac{\ell}{2}\right)\right)\frac{ \ell^2 }{4} +O(p^2\ell^2) \,,
\label{eq:Deltagrav2} 
\ee
where we are using that $|p|\ell \ll 1$. 
The discontinuity is found to be
\be
\text{Disc}_{s}[G^{\bm 2}_{{\rm AdSS}}(s)]=2\pi \delta(s)+ \frac{\pi\ell^2}{2}\,.
\ee
After substituting in Eq.\,\eqref{eq:V_spectral_gen} we obtain the gravitational potential 
\be
V_N(R)= -\frac{m_1m_2}{M^2_{\rm Pl}\,R}\left(1+\frac{2}{3}\frac{\ell^2}{ R^2}+O\left(\frac{\ell^4}{R^4}\right)\right) \,. \label{eq:VN_AdS}
\ee

The $\ell^2/R^3$ deviation is the manifestation of the continuum  $\Pi(p)$ which mixes with the 4D graviton. This is the well known behavior found in 
 \cite{Randall:1999vf}, with the exact coefficient obtained in \cite{Callin:2004py}.

\subsection{Linear dilaton}
\label{subsec:LD}

The linear dilaton (LD) model is defined by the superpotential~\cite{Megias:2021mgj}
\begin{equation}
W(\bar\phi) = \frac{6 M_5^3}{\ell} e^{\bar\phi}  \,, \label{eq:WALD}
\end{equation}
which leads to the following scalar potential
 \begin{equation}
 V(\bar\phi) = \frac{1}{6 M_5^3} \left( \frac{1}{4} \left( \frac{\partial W}{\partial \bar\phi} \right)^2 -  W(\bar\phi)^2  \right)  = - \frac{9 M_5^3}{2\ell^2} e^{2\bar\phi } \,.
 \end{equation}
This model has a solution at zero temperature which is given in conformal coordinates by
\be
\omega_{\LD}(z) = e^{- \bar\eta z } \,, \quad \bar\phi_{\LD}(z) = \bar\eta z + \log(\bar\eta \ell) \,, \label{eq:LD_sol}
\ee
with $f_{\LD}(z) = 1$, while $\bar\eta$ is a scale related to the mass gap as
\begin{equation}
\sigma = \frac{3}{2} \bar\eta \,.
\end{equation}
The solution at finite temperature is given by the same expressions of Eq.~(\ref{eq:LD_sol}), with the blackening factor
\be
f_{\LD}(z) =1 - e^{3 \bar\eta (z - z_h)} \,. 
\ee

In the brane cosmology coordinates, the black hole solution can be written as
 \begin{eqnarray}
n_{\LD}(r) &=&\frac{r}{\ell} \sqrt{1-\frac{r^3_h}{r^3} }  \,, \\
b_{\LD}(r) &=& \frac{\ell}{r_b} \frac{e^{-\bar v_b}}{ \sqrt{1-\frac{r_h^3}{r^3}} } \,, \\
\bar\phi_{\LD}(r) &=& \bar v_b -\log\left( \frac{r}{r_b} \right) \,.
 \end{eqnarray}
 We have fixed the integration constants such that the scalar vev at the brane is constant, {\it i.e.} $\bar\phi_{\LD}(r_b) = \bar v_b$. Then, the mass scale $\bar\eta$ turns out to be
\begin{equation}
\bar\eta = \eta \frac{r_{\brane}}{\ell} \,, \qquad \textrm{where} \qquad \eta \equiv \frac{e^{\bar v_b}}{\ell} \,,
\end{equation}
 and the relation between the 5D and 4D Planck scales is
\begin{equation}
M_5^3 = \eta M_{\rm Pl}^2 \,. \label{eq:M5M4_LD}
\end{equation}
 Notice that the domain of the variable $r$ is  $[0,\infty )$, where $r=0$ is the metric singularity and $r(t_0) = \ell$ is the brane location today. The black hole temperature in the LD background is
\begin{equation}
T_{\LD} = \frac{3}{4\pi} \bar\eta \,.
\end{equation}

These results are consistent with the
borderline solution  between confining and non-confining geometries reported in \cite{Gursoy:2008za}.

\subsubsection{Deviation from the Friedmann equation}

We  find the components of the effective energy density
\be
\rho^{W}_{\LD}(r_{\brane})= \frac{9}{4} \eta^2  M^2_{\rm Pl}  \frac{r_h^3}{r_b^3}
\,,\quad\quad\quad\quad  \rho^{\phi}_{\LD}(r_b) + \rho^{\Lambda}_{\LD} = 
 \frac{3}{4} \eta^2   M^2_{\rm Pl}\frac{r_h^3}{r_b^3}
\,
\ee
where the Weyl energy is  computed by using Eq.~(\ref{eq:Cr0_general}).
In this case, $\rho^{\phi}_{\LD}$ contains both a $\propto r_b^{-3}$ term and a constant term. The latter is cancelled by tuning the brane tension to $\Lambda_b = 6 \eta^2M^2_{\rm Pl}$ to set the 4D cosmological constant to zero.

The effective energy density of Eq.~(\ref{eq:rhoeff}) turns out to be
\begin{equation}
    \rho_{\rm eff,\LD} =  3 \eta^2 M_{\rm Pl}^2 
    \frac{r^3_h}{r^3_{\brane}}  \,.
    \label{eq:rhoeffLD}
\end{equation}
The  $1/r^{3}_b$ scaling amounts to $1/a^{3}$ in terms of the usual scale factor. Thus the effective energy density scales as nonrelativistic 4D matter. 

We will further discuss this interesting result in section~\ref{subsec:Discussion}, and in Ref.\,\cite{DM_LD}. 
The result passes nontrivial consistency checks. 
 We show in \cite{DM_LD} that the pressure terms arising from the $\tau^W $ and $\tau^\phi$ components of the effective stress tensor cancel out, such that the total pressure vanishes, $P_{\rm eff}=0$, consistently with the $1/a^{3}$ behavior of the total effective density. This is a consistency check ensuring that the 4D Bianchi identity is satisfied. 

 We also show in App.\,\ref{app:5D_continuity_eq}  that the conservation law Eq.\,\eqref{eq:5Dcont} is verified. This involves the nontrivial fact that 
\be
T_{MN}u^Mn^N \approx 0 \label{eq:Tun=0}
\ee
at leading order in the low energy regime.  The computation of Eq.\,\eqref{eq:Tun=0}  is detailed in App.\,\ref{app:5D_continuity_eq}.

\subsubsection{Deviation from the Newtonian potential}

In the linear dilaton background the reduced brane-to-brane graviton propagator is \cite{Megias:2021mgj}
\begin{align}
G^{\bm 2}_{{\rm LD}}(-p^2) & =-\frac{1}{2\gap^2}\frac{i}{\sqrt{1+\frac{p^2}{\gap^2}}-1} \,,
 \label{eq:Deltagrav2} 
\end{align}
where $\sigma = 3 \bar\eta / 2$ is the mass gap. This expression has both a pole at $p^2=0$ and a branch cut along  $-p^2\geq \gap^2$. The denominator can also be put in the form 
$ -p^2 +2\gap^2\left[\sqrt{1+\frac{p^2}{\gap^2}}-1-\frac{p^2}{2\gap^2}\right]  $ which reproduces the form shown in Eq.\,\eqref{eq:b2b}. The first term is the 4D pole with $m_0=0$.  The second term corresponds to the  pure continuum part which is non-analytical above $-p^2\geq \gap^2$ and $O(p^4)$ near $p\sim 0$. 
We obtain the discontinuity
\be
\text{Disc}_{s}[G^{\bm 2}_{{\rm LD}}(s)]=2\pi \delta(s) +
\frac{\sqrt{\frac{s}{\gap^2}-1}}{s}\theta\Big(s\geq\gap^2\Big) \,.
\ee
As expected the graviton spectral distribution features a massless pole and a gapped continuum. 

Substituting into Eq.\,\eqref{eq:V_spectral_gen} we obtain the gravitational potential 
\be
V_N(R)= -\frac{m_1 m_2}{M^2_{\rm Pl}\,R}\bigg(1+\Delta(R)\bigg) \,,
\ee
with
\be
\Delta(R) = \frac{2}{3\pi} \int^\infty_{\gap^2} ds \frac{\sqrt{\frac{s}{\gap^2}-1}}{s} e^{-\sqrt{s}R}\simeq \begin{cases}\frac{4}{3\pi \gap R}\qquad~{\rm if}~R\ll \frac{1}{\gap}   \\ O\Big(e^{-\gap R}\Big)~{\rm if}~R\gg \frac{1}{\gap}  
\end{cases} \,. \label{eq:VN_LD}
\ee
We see that the deviation from the Newtonian potential  appears essentially below the distance scale $1/\gap$ corresponding to the inverse mass gap. The deviation to the potential goes as $\propto 1/R^2$, unlike the AdS case where it goes as $1/R^3$.

\subsection{Asymptotically AdS linear dilaton (ALD)}
\label{subsec:ALD}

We consider a modification of the linear dilaton background featuring an AdS asymptotic behavior in the UV. The model is defined by the superpotential~\cite{Cabrer:2009we,Megias:2019vdb}
\begin{equation}
W(\bar\phi) = \frac{6 M_5^3}{\ell} \left( 1 + e^{\bar\phi} \right) \,, \label{eq:WALD}
\end{equation}
which leads to the following scalar potential
\begin{equation}
V(\bar\phi) = - \frac{6 M_5^3}{\ell^2} \left( 1 + 2 e^{\bar\phi} + \frac{3}{4} e^{2\bar\phi} \right) \,.
\end{equation}
The metric we are considering is, using proper and brane cosmology coordinates,
\begin{eqnarray}
ds^2 &=& e^{-2A(y)} \left( -h(y) d\tau^2 + d\x^2  \right) + \frac{dy^2}{h(y)} \,, \\
 &=& - n(r)^2 d\tau^2 + \frac{r^2}{\ell^2} d\x^2 + b(r)^2 dr^2 \,. \label{eq:ds2r}
\end{eqnarray}

The solution of the background equation of motion, in proper coordinates, is
\begin{eqnarray}
A_\ALD(y) &=& \frac{y}{\ell} - \log\left( 1 - \frac{y}{y_s}\right) \,, \\
h_\ALD(y) &=& 1 - \frac{ \int_{-\infty}^y d{\bar y} \, e^{4A_{\ALD}({\bar y})} }{ \int_{-\infty}^{y_h} d{\bar y} \, e^{4A_{\ALD}({\bar y})} } \,, \\
\bar\phi_\ALD(y) &=& -\log\left( \frac{y_s - y}{\ell} \right) \,,   \label{eq:ALD_bkg}
\end{eqnarray}
where $y_s$ is the location of the \textit{naked} singularity, which would correspond to $z_s\to\infty$ in conformal coordinates. In the brane cosmology coordinates, the solution is given by~\footnote{The relation between the proper ``$y$'' and the brane cosmology ``$r$'' coordinates in the ALD model of Sec.~\ref{subsec:ALD} is
\begin{equation}
\frac{r}{\ell} = \left( 1 - \frac{y}{y_s} \right) e^{-y/\ell}  \qquad \textrm{or its inverse} \qquad  y = y_s \left( 1 -  \frac{\mathcal W( cr/r_b)}{\mathcal W(c\ell/r_b)} \right) \,, \label{eq:ry_inv}
\end{equation}
where $\mathcal W(z)$ is the principal branch of the Lambert function.
}
\begin{eqnarray}
    n_\ALD(r) &=& \frac{r}{\ell} \sqrt{h_{\ALD}(y(r))} \,,  \\
    b_\ALD(r) &=& \frac{1}{\sqrt{h_{\ALD}(y(r))}} \frac{\ell}{r} \frac{\mathcal W\left( c \frac{r}{r_b} \right) }{1 + \mathcal W\left( c \frac{r}{r_b} \right) }   \,,  \\
    \bar\phi_\ALD(r) &=& -\log \mathcal W\left( c \frac{r}{r_b}  \right)   \,, 
\end{eqnarray}
with
\begin{equation}
    c = e^{-\bar v_b + e^{-\bar v_b}} \,,
\end{equation}
and $\mathcal W(z)$ is the principal branch of the Lambert function. 

As in the LD model of Sec.~\ref{subsec:LD}, in the ALD model the graviton spectrum has a mass gap $\sigma = 3\bar\eta/2$, with
\begin{equation}
\bar\eta = \frac{1}{y_s} e^{-y_s/\ell} = \eta \frac{r_{\brane}}{\ell} \,, \qquad \textrm{where} \qquad \eta \equiv \frac{1}{c\ell} \,.
\end{equation}
Moreover, the relation between the 5D and 4D Planck scales is
\begin{equation}
M_5^3 =  \frac{M_{\rm Pl}^2}{\ell} \left( 1 + e^{\bar v_{\brane}} \right) \,. \label{eq:M5M4_LDA}
\end{equation}
\subsubsection{Deviation from the Friedmann equation}

We find the components of the effective energy density
\begin{equation}
\rho^{W}_{\ALD}(r_{\brane}) =  \frac{3M^2_{\rm Pl}}{4\ell^2} \frac{\left( 1 + e^{\bar v_b} \right) }{ c^4 I\left( c\, r_h / r_b \right)} \,, \qquad \rho^{\phi}_{\ALD}(r_b) + \rho^{\Lambda}_{\ALD} =  \frac{3 M^2_{\rm Pl}}{4\ell^2} e^{2\bar v_b} \frac{I(c)}{I\left( c\, r_h / r_b \right)}   \,,  \label{eq:rhoWphi_ALD}
\end{equation}
where the function $I(\chi)$ is defined in Eq.~(\ref{eq:I_chi}).  More details can be found in App.\,\ref{app:Technical_details}. The Weyl energy is  computed by using Eq.~(\ref{eq:Cr0_general}). As in the LD model, $\rho^{\phi}_{\ALD}(r_b)$ contains a constant term which is cancelled by tuning the brane tension to $\Lambda_{\brane} = \frac{6}{\ell^2}\left( 1 + e^{\bar v_{\brane}}\right)^2 M_{\rm Pl}^2$ to set the 4D cosmological constant to zero.

In the ALD background, the conservation equation~(\ref{eq:5Dcont}) can be solved in the asymptotic limits $c \to 0$ and $c \to \infty$. If $c\ll 1$, {\it i.e.}~$\bar v_b\gg 1$, since $r_b>r_h$ we are always in the regime $r_b\gg c\, r_h$.  Thus the effective energy density is Eq.~(\ref{eq:rhoeffLD}) just like  in the pure linear dilaton background, and $\rho_{\rm eff, ALD}$ behaves as nonrelativistic matter. On the other hand,  if  $c\gg 1$, \textit{i.e.} $\eta\ell\ll1$ (\textit{i.e.} $\bar v_b<0,\ |\bar v_b|\gtrsim 1$), we find that $\rho^\phi_{\ALD}(r_b) + \rho^\Lambda_{\ALD} \sim c^{-2} \rho^W_{\ALD}(r_b)$ so that the Weyl energy is dominant in this region. The total effective energy density of Eq.~(\ref{eq:rhoeff})  for $c \gg 1$ turns out to be
\begin{equation}
   \rho_{\rm eff,\ALD}(r_b) \simeq \rho^W_{\ALD}(r_b)  \simeq  \frac{3}{\ell^2} M_{\rm Pl}^2 \frac{r_h^4}{r_b^4}  \qquad \qquad (c \gg 1) \,.
\end{equation}
In this limit, $\rho_{\rm eff, ALD}$ behaves as radiation just like in pure  AdS background.

\begin{figure}[t]
\centering
\includegraphics[width=0.45\textwidth]{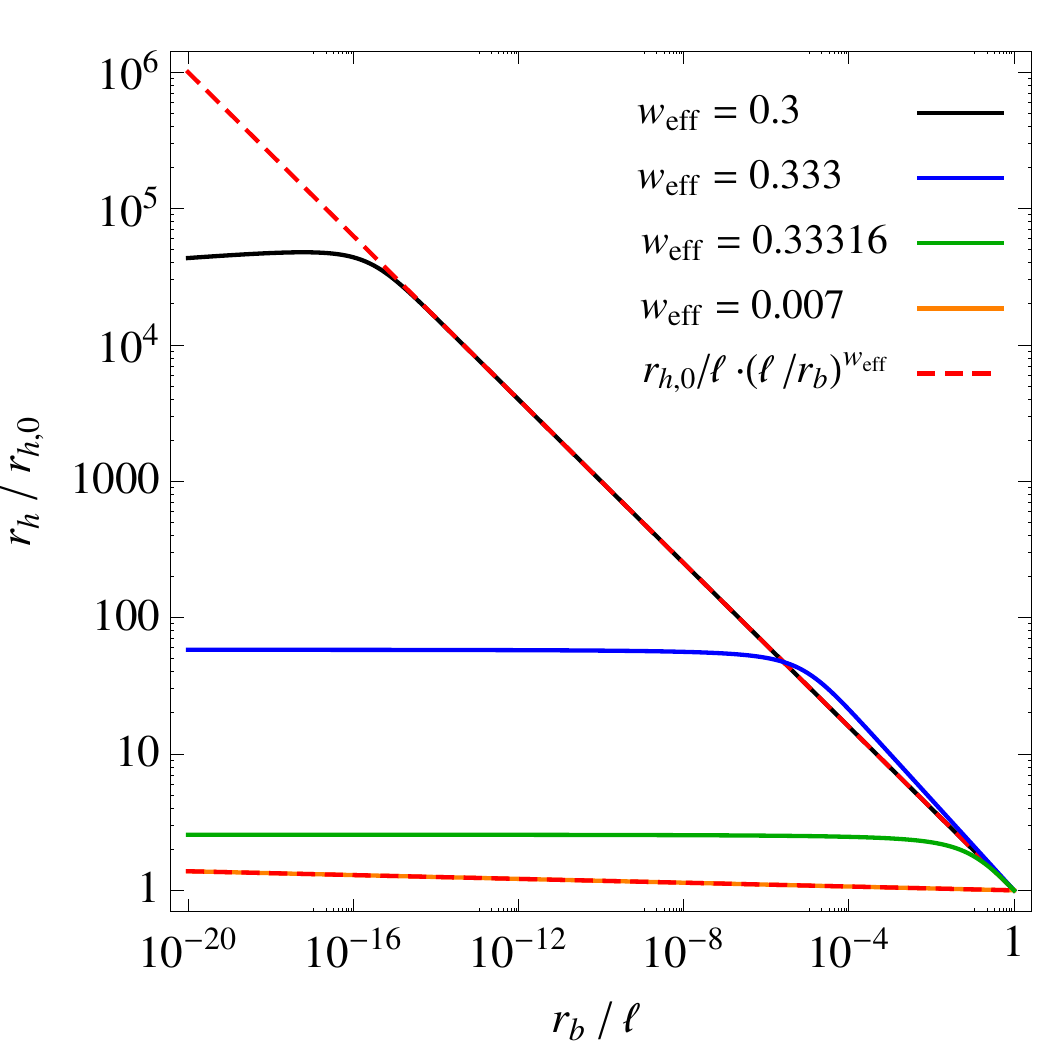}
\hspace{0.5cm}
\includegraphics[width=0.45\textwidth]{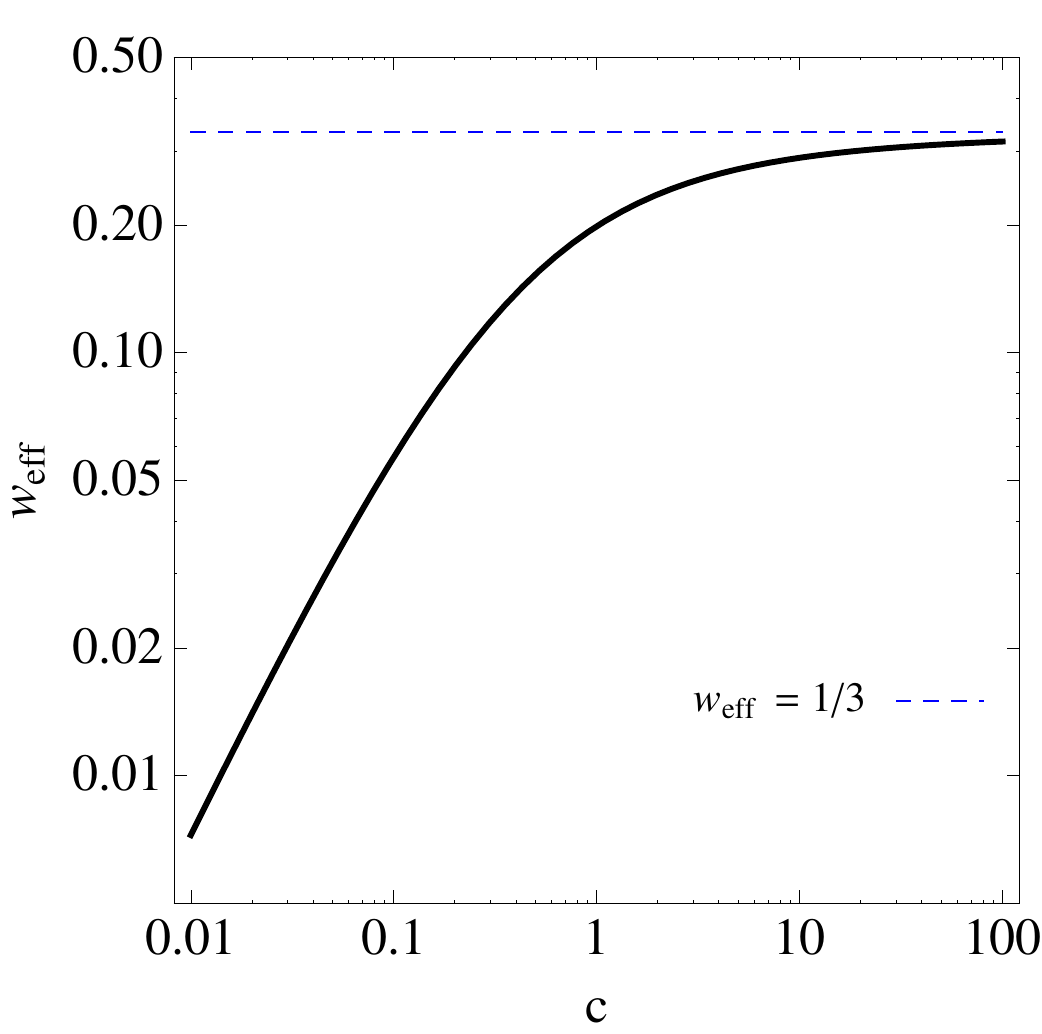} 
\caption{\it Results within the asymptotically-AdS linear dilaton model. Left panel: Location of the black hole horizon ($r_h$) as a function of the brane location ($r_b$)  enforced by the conservation equation~(\ref{eq:5Dcont}). We display as dashed red lines the analytical results of Eq.~(\ref{eq:rhrb_analytic}) for $r_{\brane}^\ast < r_{\brane}$. We use $r_{h,0}/\ell = 10^{-22}$. Right panel: Equation-of-state parameter $w_{\textrm{eff}} \equiv P_{\rm eff} / \rho_{\rm eff}$ as a function of~$c$. To guide the eye, we display the asymptotic value $w_{\textrm{eff}} = 1/3$ (dashed blue line).}
\label{fig:w}
\end{figure}

For arbitrary values of $c$, the solution of the conservation equation,  Eq.~(\ref{eq:5Dcont}), demands  
that the black hole horizon depends on the brane location,
\textit{i.e.}~$r_h = r_h(r_b)$. We display in the left panel of Fig.~\ref{fig:w} the dependence $r_h(r_b)$ for various values of the parameter $w_{\rm eff}$. One gets the analytical behavior~\footnote{In Eq.~(\ref{eq:rhrb_analytic}) we have used the following definitions 
\begin{equation}
\frac{r_{\brane}^\ast}{\ell} \equiv \left( \frac{4c}{3} \frac{r_{h,0}}{\ell} \right)^{1/(1+w_{\rm eff})} \qquad \textrm{and} \qquad   \frac{\beta_{\ast}}{\ell} \equiv  \left( \frac{3}{4c}\frac{r^3_{h,0}}{\ell^3} \right)^{1/4}     \,.
\end{equation}
The expression of $\beta_{\ast}$ ensures continuity of $r_h(r_{\brane})$ at $r_{\brane} = r_{\brane}^\ast$. Notice that the condition $c < \ell/r_{h,0}$ is approximately equivalent to $r_{\brane}^\ast/\ell < 1$. Notice also that in the regime $c \ll 1$ one has $r_{\brane}^\ast \simeq c\, r_{h,0} \ll r_{h,0}$, and so in this case $r_{\brane}^\ast \ll r_{h,0} \le r_{\brane}$.}
\begin{equation}
r_h(r_b) \simeq 
   \begin{cases}
\beta_\ast \left( \frac{\ell}{r_b} \right)^{(3 w_{\rm eff} - 1)/4} 
 \,\;\quad\;\,\,  {\rm if}  ~~  r_{\brane} < r_{\brane}^\ast
 \\
r_{h,0} \left( \frac{\ell}{r_b} \right)^{w_{\rm eff}} \,\;\quad\;\;\qquad\; {\rm if} ~~ r_{\brane}^\ast< r_{\brane} 
   \end{cases}   \qquad \left( c < \ell/r_{h,0} \right)  \,, \label{eq:rhrb_analytic}
\end{equation}
 while $r_h(r_{\brane}) \simeq r_{h,0} \left( \ell/r_b \right)^{(3 w_{\rm eff} - 1)/4} \simeq r_{h,0}$ is obtained at $\ell/r_{h,0} < c$. In these formulas,  $w_{\rm eff}(c) = P_{\rm eff} / \rho_{\rm eff}$ is the equation-of-state parameter, and $r_{h,0}$ is the radius of the black hole horizon at present times $t = t_0$. Finally, the effective energy density turns out to have the following behaviors~\footnote{In writing the expression in Eq.~(\ref{eq:rhoeffALDrb}) for the regime $1 \ll c < \ell/r_{h,0}$ we have assumed that $r_{\brane}^\ast < r_{\brane}$. The corresponding expression for $r_{\brane} < r_{\brane}^\ast$ in that regime is  $\rho_{\rm eff,\ALD}(r_b) \simeq \frac{3}{\ell^2} M_{\rm Pl}^2 \frac{r_h^4(r_b)}{r_b^4}$, which after using that $r_h(r_{\brane}) \simeq \beta_\ast$ leads to the same behavior as the one shown in the rhs of Eq.~(\ref{eq:rhoeffALDrb}).}
\begin{equation}
\hspace{-0.04cm}   \rho_{\rm eff,\ALD}(r_b) \simeq 
   \begin{cases}
   3 \eta^2 M_{\rm Pl}^2 \frac{r_h^3(r_b)}{r_b^3}  \simeq  3 \eta^2 M_{\rm Pl}^2 
    \frac{r_{h,0}^3}{r^3_{\brane}} 
 \,\;\quad\quad\;\;  {\rm if} ~~ c \ll 1 
 \\
\frac{9}{4\ell} \eta M_{\rm Pl}^2 \frac{r_h^3(r_b)}{r_b^3} \simeq \frac{9}{4} \eta M_{\rm Pl}^2 \frac{r_{h,0}^3}{r_b^4} \,\;\quad\;\;\quad\; {\rm if} ~~ 1 \ll c < \ell/r_{h,0} 
 \\
 \frac{3}{\ell^2} M_{\rm Pl}^2 \frac{r_h^4(r_b)}{r_b^4} \simeq \frac{3}{\ell^2} M_{\rm Pl}^2\, \frac{r_{h,0}^4}{r_b^4}  \quad\quad\qquad {\rm if} ~~  1 \ll \ell/r_{h,0} < c
   \end{cases}   \,.   \label{eq:rhoeffALDrb}
\end{equation}
The equation of state smoothly interpolates between matter and radiation behavior, $\rho_{\rm eff,\ALD}\propto a^{-3(1+w_{\rm eff}(c))}$. The numerical value of $w_{\rm eff}(c)$ is exhibited in the right panel of Fig.~\ref{fig:w} where a continuous transition appears between $w_{\rm eff} = 0$ (matter) and $w_{\rm eff} = 1/3$ (radiation). We provide in Appendix~\ref{subsec:app_ALD} the explicit exact analytical expressions of $\rho_{\rm eff,\ALD}(r_b)$ and $w_{\rm eff}(c)$, cf.~Eqs.~(\ref{eq:rhoeff_rhrb})-(\ref{eq:weff_analytic}). We confirm all these results via numerical solving of the 5D conservation equation~(\ref{eq:5Dcont}).

\subsubsection{Deviation from the Newtonian potential}
\label{subsec:Newton_ALD}

The  equations of motion for the graviton propagator on the ALD background do not have exact analytical solutions. However an approximation is easily obtained by considering two regimes. The metric is approximately AdS for $z\ll 1/\sigma$ and LD for $z\gg 1/\sigma$. 
On the other hand, at the level of propagation we know that AdS propagators, expressed in $(p_\mu,z)$ space with given spacelike momentum $p_\mu$,  are exponentially suppressed beyond $z\sim 1/p$ (see \textit{e.g.}~\cite{Randall:2001gb, Costantino:2020vdu}). That is, the propagator only knows about the  $z \lesssim 1/\sigma$ region of the bulk.  This fact implies that if $\sqrt{s} \gg \sigma$ the spectral function should not know about the LD part of the background, and thus be approximately AdS. On the other hand, for $\sqrt{s}\ll\sigma$ the propagator should know about the LD background. But since the LD background induces a mass gap at $\sigma$, the dominance of the LD background implies that the continuum vanishes. This is consistent with the spectral function obtained in our approximation, in which the continuum part  starts at $\sqrt{s} = \sigma$.

\begin{figure*}[t]
\centering
\includegraphics[width=0.40\textwidth]{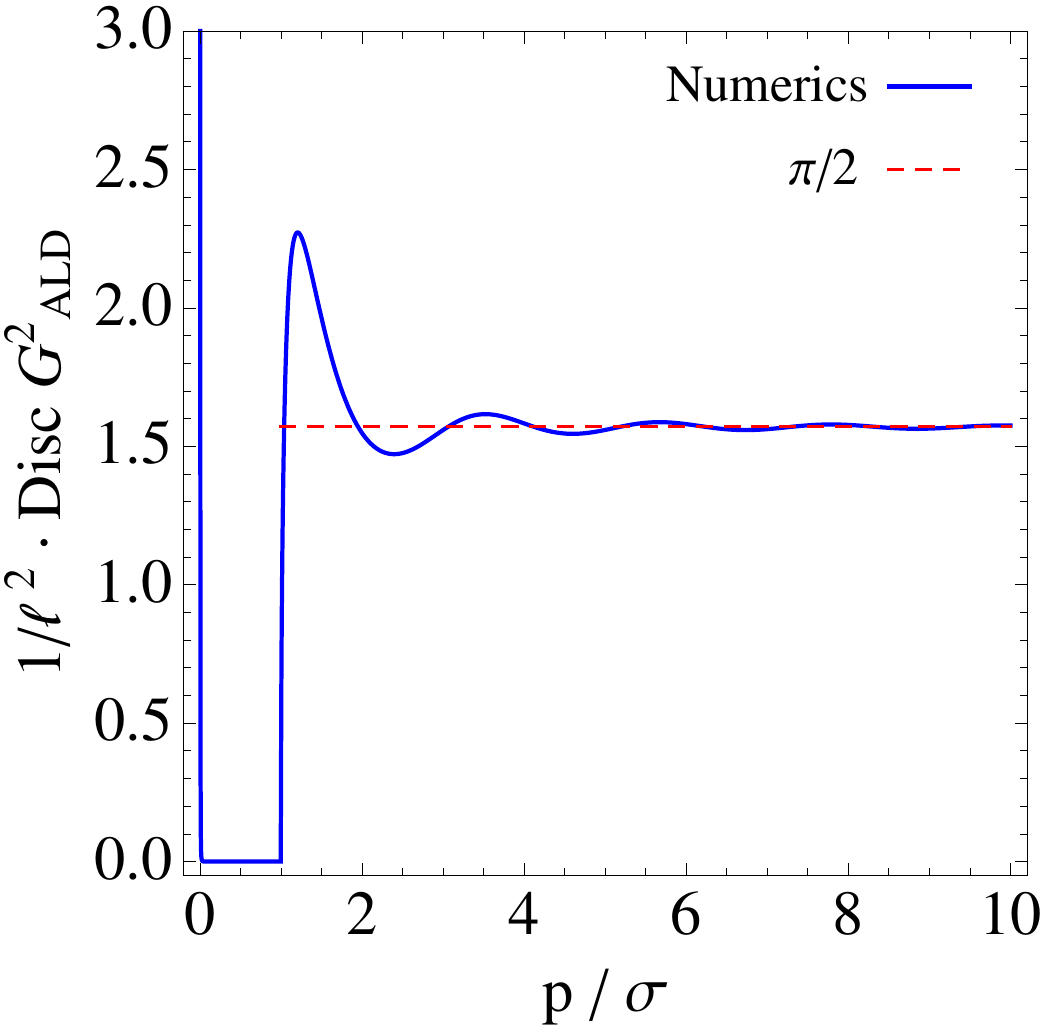}  \hspace{1cm} 
\includegraphics[width=0.42\textwidth]{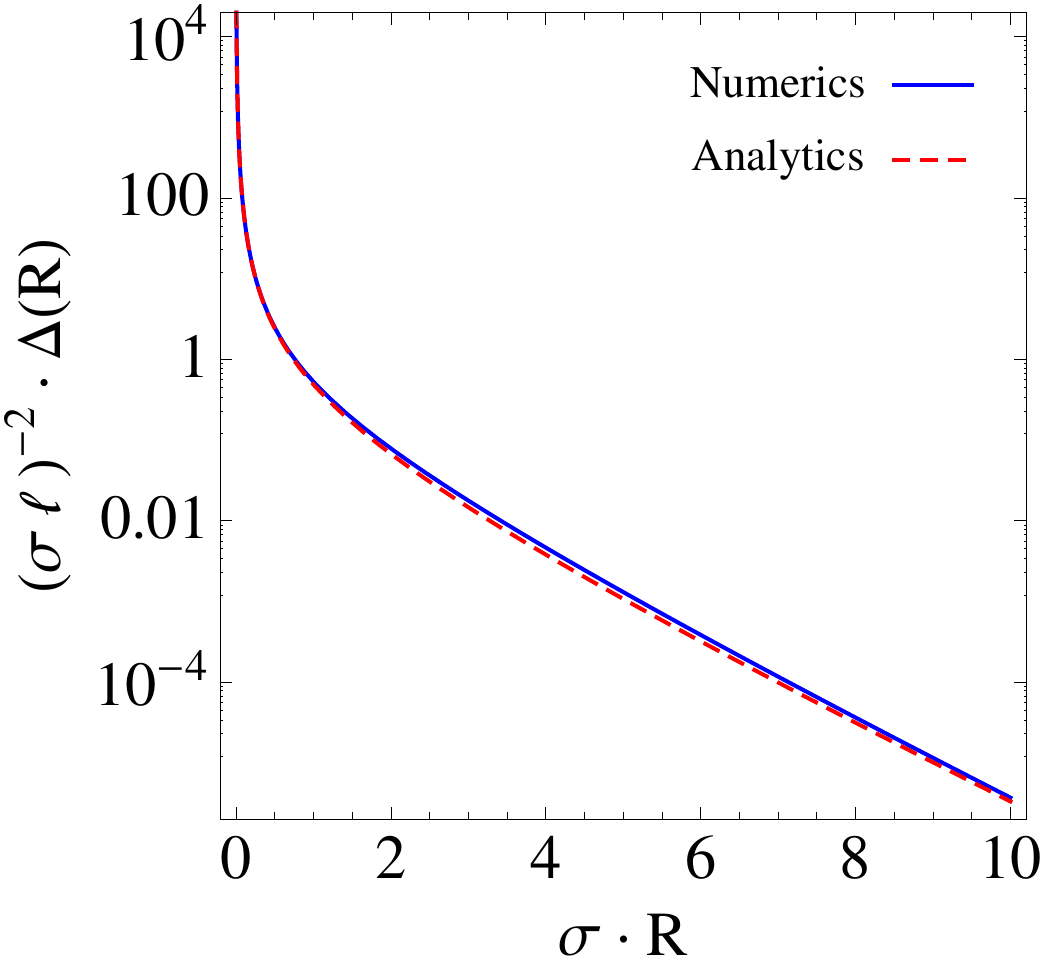} 
 \caption{\it Results within the ALD model of Sec.~\ref{subsec:ALD}. We display the discontinuity of the graviton propagator (left panel), and the deviation from the Newtonian potential (right panel). The analytical result displayed in the right panel corresponds to the correction term inside the bracket in Eq.~(\ref{eq:VN_ALD}). In this plot we have considered a piecewise approximation of the ALD model.}
\label{fig:VN}
\end{figure*}

In summary we can  approximate the discontinuity of the graviton propagator as 
\be
\text{Disc}_{s}[G^{\bm 2}_{{\ALD}}(s)]  \approx 2\pi \delta(s) +
\frac{\pi \ell^2}{2} \theta\Big(s\geq\gap^2\Big)  \,.
\label{eq:DiscALD}
\ee
The Newtonian potential is easily computed by plugging Eq.\,\eqref{eq:DiscALD} into Eq.\,\eqref{eq:V_spectral_gen}, giving
\be
V_N(R)\approx -\frac{m_1m_2}{M^2_{\rm Pl}\,R}\left(1+
\frac{2\ell^2}{3 R^2} e^{-\sigma R}(1+\sigma R)
\right) \,.  \label{eq:VN_ALD}
\ee
We can see that for $R\ll 1/\sigma$ the expression reduces to the AdS one, Eq.\,\eqref{eq:VN_AdS}. On the other hand for $R> 1/\sigma$ the potential is exponentially suppressed --- as a consequence of the mass gap induced by the LD background. We also evaluate numerically in Fig.\,\ref{fig:VN} the results of $\text{Disc}_{s}[G^{\bm 2}_{{\ALD}}(s)]$ and $V_N(R)$ by considering the piecewise approximation of the background solution of Eq.~(\ref{eq:ALD_bkg}), cf.~Ref.~\cite{Megias:2021arn}. Nontrivial oscillations occur near the threshold that cannot be captured analytically. Despite this detail the numerical evaluation of the potential accurately reproduces the analytical behavior.

\subsection{Discussion}
\label{subsec:Discussion}

Overall we have found that the deviations from the Newtonian potential and Friedmann equation appearing in the linear dilaton background completely differ from those occurring in the AdS background.

First, the deviation from the Newtonian potential induced in the linear dilaton background goes as $1/(\sigma R^2)$  and is gapped at $R\sim 1/\sigma$. In contrast, the deviation from gravity in the AdS background goes as $\ell^2/R^3$ and is ungapped. We can use the landscape of Yukawa-like fifth force searches to bound the deviation.
We find that the relevant bound is the one from micron-scale fifth force experiments~\cite{Smullin:2005iv}. The order of magnitude bound is
\be
\frac{1}{\sigma}\lesssim 10~ \mu{\rm m} \label{eq:sigma_bound}
\ee
or $\sigma\gtrsim 0.01$~eV.

Second, in the brane-world paradigm, the energy of standard matter is identified with $\rho_b$. Therefore the $\rho_{\rm eff}$ energy density emerging in the linear dilaton background can be identified as \textit{dark matter}. This energy density scales as $a^{-3}$, or equivalently has vanishing pressure such that its equation of state parameter is 
\be
\omega_{\rm eff, LD}=0\,. 
\ee
Notice that in AdS-Schwarschild background one has instead $\omega_{\rm eff, AdSS}=\frac{1}{3} $, \textit{i.e.}~dark radiation.  

The dark matter density predicted from our linear dilaton brane-world
can be translated into the  dark matter  density parameter $\Omega_{{\rm DM}}= \rho_{\rm eff}/\rho_{\rm crit}$  with  the critical energy density  $\rho_{\rm crit}= 3H^2 M^2_{\rm Pl}$. The result is 
\be
\Omega_{\rm DM,\,LD} = \left(\frac{\eta}{H}\right)^2 \left(\frac{r_h}{r_b}\right)^3\,.
\label{eq:OmegaDMLD}
\ee
At present times $t=t_0$, we have $r_b(t_0)=\ell$ (\textit{i.e.} $a(t_0)=1$) and we know that $\Omega_{{\rm DM},0} = 0.267$ and $H_0=1.47 \times 10^{-42}$\,GeV. 
This imposes a relation between the parameters of the linear dilaton background.  Combined with the fifth force bound Eq.\,\eqref{eq:sigma_bound}, one can also verify that $r_{h,0}< 2.3 \cdot 10^{-21} \ell$.

In summary, a brane-world living in the linear dilaton background automatically contains dark matter. We further expand on this remarkable feature in \cite{DM_LD}.

Finally, the ALD background --- a blend  of the AdS and linear dilaton backgrounds --- is interesting for both conceptual and phenomenological reasons. 
It is conceptually instructive because it teaches us more about the properties of the holographic theory. Namely, since the bulk is asymptotically either AdS or linear dilaton, one can wonder which  regime appears from the brane viewpoint.
 Regarding the deviation from the Newtonian potential, we found that the AdS regime emerges in the UV, \textit{i.e.}~for small $R$, while the LD regime shows up in the IR, \textit{i.e.}~for large $R$. 
Thus, for this observable, both regimes show up upon variation of a physical parameter. 
The situation of the Friedmann equation is  different: the scaling of $\rho_{\rm eff}$ only depends on the value of the scalar vev on the brane. Thus the behavior of $\rho_{\rm eff}$   does not vary as a function of a physical parameter such as time or temperature.   From this we conclude that the manifestation of  the AdS and linear dilaton regimes to a brane observer is subtle, in the sense that it depends on the observable considered.

From a phenomenological viewpoint, the fact that the ALD model provides an energy density with $0\leq w_{\rm eff}\leq \frac{1}{3}$ is an interesting feature. It turns out that considering the dark matter equation of state as a free parameter has been done in the framework of ``generalized dark matter'', see \cite{Hu_GDM,Kopp:2018zxp,Thomas:2019dds}. Our ALD model can therefore be taken as a UV completion of this framework. 
In the $c\ll 1$ limit we have $ w_{\rm eff}\to 0$, in which case one can identify $\rho_{\rm eff}$ as the dark matter and use Eq.\,\eqref{eq:OmegaDMLD}. In the $c\gg 1$ limit we have 
$ w_{\rm eff}\to \frac{1}{3}$. In this limit the predicted fraction of dark radiation   reads
\be
\Omega_{\rm DR,\,ALD} \bigg|_{c\gg 1} = \Omega_{\rm DR,\,AdSS}  = \frac{1}{ \left(\ell H\right)^2} \left(\frac{r_h}{r_b}\right)^4\,.
\label{eq:OmegaDMALD}
\ee
Using that $\Omega_{\rm DR}\approx 0.135 \Delta N_{\rm eff}\lesssim 0.07$ at BBN times\,\footnote{
Assuming standard thermal history the bound on the effective number of neutrinos $N_{\rm eff}\approx 3+\Delta N_{\rm eff}$
translates into a bound on the fraction of dark radiation as
\be \Omega_{\rm DR}= \frac{\frac{7}{4}\left(\frac{4}{11}\right)^{\frac{4}{3}}}{2+\frac{21}{4}\left(\frac{4}{11}\right)^{\frac{4}{3}}} \Delta N_{\rm eff}+O\left((\Delta N_{\rm eff})^2\right)\,.
\ee 
A typical bound from BBN is   $\Delta N_{\rm eff} \lesssim 0.5$ \cite{Lin:2019uvt}. 
}  and combining  with the fifth force bound $\ell \lesssim 5 \,\mu$m (obtained by using the bounds in \cite{Brax:2017xho} on the potential Eq.\,\eqref{eq:VN_AdS}), we can also verify that $r_{h,0} \lesssim  10^{-16} \ell$. This bound is weaker  than for the linear dilaton model.

 The cosmological brane-world scenarios presented here  certainly deserve further investigation.

\section{Summary }

Here we summarize the logical steps and results of our study. 

 The first part of the paper is a broad  analysis of continuum EFTs.
Our interest lies in theories giving rise to a free continuum in some parametric limit. A theory featuring a free continuum is referred to as a GFT. In our definition of GFT we allow for local interactions of the continuum, as this has no impact on the results. A free continuum sector emerges in the limit of theories with nontrivial dynamics, such as the $N\to\infty$ limit of gauge theories or the $g_{\rm bulk}\to 0$ limit of 5D holographic models. Additionally, a GFT may be seen as an approximation of a discretum.  Standard Poincar\'e-invariant (\textit{i.e.}~no brane) weakly coupled QFTs do not give rise to a continuum in the free limit, thus these  are excluded from our study. 

There is a priori no obvious principle to prevent us from writing an EFT featuring a free continuum. However we argue that such an EFT is incompatible with standard gravity. One line of argument is to show that the continuum sector has no stress tensor, or that the central charge is infinite. An axiomatic version of this fact is known for CFT and reviewed here. Using the continuous mass representation we obtain a similar conclusion for any non-conformal free continuum. 
Another line of reasoning relies on the species scale of gravity. The species scale is usually given for stable particles. Here, as a side result, we present a finite-temperature-based argument that generalizes the species scale in terms of the central charge of any CFT. Using the species scale we argue that the free continuum sector amounts to an infinite number of species, and thus that the cutoff of the EFT is zero. 

These arguments imply that a free continuum in the presence of standard gravity cannot exist. We then consider the neighborhood of this point in theory space, that evade the no-go arguments, either because  the number of degrees of freedom is finite or gravity is nonstandard. 
This is the case of the classes of theories already listed above: a discretum,  gauge theories with finite $N$, holographic theories. 
We point out that a common feature of all these models  is that they must feature significant deviations in the gravity sector --- these are the effects blowing up when approaching the GFT+4D Einstein gravity point.

 Guided by this general analysis, in the second part of the paper we focus on holographic theories giving rise to a continuum. We consider a class of 5D scalar-gravity models that gives rise to a gapped continuum.  We lay out --- together with a review of QFT aspects needed for an overall understanding of the holographic framework --- the necessary formalism to compute the Newtonian potential and the effective Friedmann equation. When brane-localized fields are at finite temperature, a horizon forms in the bulk.
We solve analytically the pure linear dilaton background at finite temperature. We also introduce a simple modification of the bulk potential which makes the background interpolating between AdS (in the UV) and linear dilaton (in the IR). We compute  this asymptotically linear dilaton background at finite temperature using both analytical approximations and exact numerical solving. 
    
In the pure linear dilaton background, we find that the Newtonian potential features a $\sim 1/(\sigma R^2)$ deviation and has a mass gap at $R\sim 1/\sigma$. This is in sharp contrast with the deviation in the AdS background. 
At finite temperature there is a horizon in the bulk. We find that the effective Friedmann equation features a holographically-induced energy density   with $a^{-3}$ scaling, due to contributions from both Weyl tensor and bulk stress tensor. Summarizing, 
\be
\boxed{
\textrm{Linear~dilaton~horizon} ~~ \Longleftrightarrow ~~ w_{\rm eff}= 0  
}
\ee
in terms of the equation of state parameter.  
Interpreting the setup as a brane-world where standard matter is brane-localized, the effective energy density is identified as \textit{dark matter}. 
In short, a brane-world living in the linear dilaton background automatically contains dark matter.  %
This is, again, in contrast with the AdS brane-world  for which the holographically-induced energy density scales as dark radiation, $w_{\rm eff}=\frac{1}{3}$. 

We also study a somewhat more evolved linear dilaton background with AdS asymptotics near the boundary. The Newtonian potential is found to be essentially like the AdS one, but with a gap at $R=1/\sigma$ like in the LD case. The behavior of the effective energy density depends on the value of the bulk scalar vev on the brane.  The equation of state  smoothly interpolates between cold dark matter  and dark radiation, \textit{i.e.}~$0 \leq w_{\rm eff}\leq\frac{1}{3}$, as a function of the scalar vev. { As a  nontrivial check of our results,  we verify that the conservation law  of the effective energy density is always satisfied. This happens via cancellations  involving the brane kinematics.  }

A general lesson from our study of the holographic models is that the cosmology of continuum models is highly nontrivial. This is, in a sense,  because it necessarily involves the underlying dynamics giving rise to the continuum. 
Here we have studied a particular case of the scalar-gravity system. A host of solutions remains to be explored. The cosmological history of the associated brane-world models certainly deserves deeper investigation. These exciting directions are left for future work.

\label{se:conclusion}

\section*{Acknowledgments}

The authors thank Philippe Brax, Csaba Csaki and Flip Tanedo for useful discussions and comments. The work of SF has been supported by the S\~ao Paulo Research Foundation (FAPESP) under grants \#2011/11973, \#2014/21477-2 and \#2018/11721-4, by CAPES under grant \#88887.194785, and by the University of California, Riverside. EM would like to thank the ICTP South American Institute for Fundamental Research (SAIFR), S\~ao Paulo, Brazil, for hospitality and partial finantial support of FAPESP Grant  2016/01343-7 from Aug-Sep 2022 where part of this work was done. The work of EM is supported by the project PID2020-114767GB-I00 funded by MCIN/AEI/10.13039/501100011\-033, by the FEDER/Junta de Andaluc\'{\i}a-Consejer\'{\i}a de Econom\'{\i}a y Conocimiento 2014-2020 Operational Programme under Grant A-FQM-178-UGR18, and by Junta de Andaluc\'{\i}a under Grant FQM-225. The research of EM is also supported by the Ram\'on y Cajal Program of the Spanish MICIN under Grant RYC-2016-20678. The work of MQ is partly supported by Spanish MICIN under Grant FPA2017-88915-P, and by the Catalan Government under Grant 2017SGR1069. IFAE is partially funded by the CERCA program of the Generalitat de Catalunya.

\appendix

\section{On the transition between discretum and continuum }
\label{app:discretum}

In this appendix we expand on the possiblity of a discretum EFT, studying its validity range using general arguments. 

For concreteness we  assume that the discretum arises as the low-energy limit of a confining large $N$ Yang-Mills theory with large `t Hooft coupling $\lambda\gg1$. 
We choose the spectral distribution of the free propagator to be
\be
\rho^{\rm free}_{\rm d}(s) = \sum^\infty_{n=n_0}  \rho_n \delta(s-s_n)  \,,
\ee
where the $s_n$ are intervals with some spacing $\step^2$ set by some typical scale $\step$. 
The sum starts at $n_0$, with $s_{n_0}=\sigma^2$. We refer to $\step$ as the mode spacing and $\sigma$ as the mass gap of the spectrum. The gap $\sigma$ is either $0$ or $\geq \step$. In the following we assume $\sigma\gg \step$. The conclusions are trivially extended to the cases $\sigma=0$ and $\sigma=O(\step)$ by replacing $\sigma$ by $\step$ in the arguments.

The free propagator takes the form
\be
 \langle{\cal O}(p){\cal O}(-p)\rangle_{\rm free} = i \sum^\infty_{n_0}  \frac{\rho_n }{-p^2-s_n+i\epsilon}\, \label{eq:2pt_GFT_disc}\,.
 \ee
It encodes a series of free 4D particles.  Similarly to Sec.\,\ref{se:CMR} the model is equivalently written with a set of canonically normalized 4D fields $\{\varphi_{n}\}$ with $\varphi_{s_n}\equiv \varphi_n$ and the operator ${\cal O}=  \sum^\infty_{n=n_0}  \sqrt{\rho_n} \varphi_n$.  

A similar picture is also obtained from holographic models with a discrete spectrum~\cite{Costantino:2020msc}. 
In the context of phenomenological continuum models, some aspects of the discretum EFT were  discussed in Ref.~\cite{Stephanov:2007ry}. 

Assuming that the discretum arises from confinement of gauge theory, the $\varphi_n $ fields can be understood as glueball fields. The couplings among the $\varphi_n$ fields are then controlled by powers of  $1/N$~\cite{Witten:1979kh}. The modes encoded into the full 2pt function are thus narrow --- in accordance with our definition of discretum. 
The  discretum EFT has a validity cutoff scale $\tscale$, above which the gauge theory description takes over, and above which ${\cal O}(p){\cal O}(-p)$ is  a continuum.  
This means that in the spectral distribution of the 2pt function there should be a transition, between the discrete and the continuous regimes, as a function of the squared mass variable~$s$.
What can we learn from general considerations about the transition scale $\tscale$?
 
We can reason in terms of degrees of freedom. On very general grounds, the number of degrees of freedom should decrease when the RG flow goes toward the IR. The UV theory (\textit{i.e.}~the deconfined gauge theory) has $\sim N^2$ degrees of freedom. Hence the low-energy effective theory can have at most $\sim N^2$ degrees of freedom. Hence the heaviest field of the EFT has at most a squared mass of $s_{n_0+N^2}$.
Moreover, since the $\varphi_n$ fields are by assumption regularly spaced, the cutoff has to be of order of the heaviest field of the discretum EFT in order to truncate the heavier ones.
 We conclude that the transition scale is constrained to be 
 \be \tscale \lesssim \sqrt{s_{n_0+N^2}} \,.\ee

We can also reason in terms of interactions. Using large-$N$ counting rules for glueballs, the 3pt interaction of the $\varphi_n$ fields has $1/N$ strength and we can then evaluate the width of an individual field $\varphi_n$. A very rough estimate is $\Gamma_n\sim \sqrt{s_n} (n-n_0) /N^2$, where $n-n_0$  counts the lighter states available for decay. Therefore the $\varphi_n$ fields become broad (\textit{i.e.}~$\Gamma_n\sim \sqrt{s_n}$) at $n\sim n_0+ N^2$, which signals a breakdown of the EFT.
We conclude that the cutoff cannot be higher than $\sqrt{s_{n_0+N^2}}$. 
This estimate matches the one from the number of degrees of freedom.

A more refined estimate can also be obtained using input from holographic models, and in particular, and for simplicity, using a pure AdS two-branes model (see Ref.~\cite{Costantino:2020msc}). In that case the spacing is $s_n=n^2\step^2$, we have $\sigma=0$, and we know that the width estimate is rather $\Gamma_n\sim \sqrt{s_n} /N^2$ because the selection rules set by the residual symmetries constrain the decay channels. We also know that the transition scale is reached when 
 the modes tend to overlap with each other, in which case not only the diagonal width $\Gamma_n$, but the full self-energy matrix that mixes all the $\varphi_n$ would become relevant~\cite{Fichet:2019hkg,Costantino:2020msc}. At that scale the $\varphi_n$'s merge, giving rise to a continuum.  The estimate of the transition scale in this case is  $\tscale \sim \sqrt{s_{N^2}}=N^2\step $. This matches the previous one when using $\sigma=0$.

 \section{Solutions of the 5D scalar-gravity system}
\label{app:Technical_details}

We present in this appendix the most general solutions of Eqs.~(\ref{eq:EoM1})-(\ref{eq:EoM4}), both in  conformal coordinates  and in brane cosmology coordinates.  The relations among integration constants are also discussed.  

\subsection{Coordinate systems }
We use three coordinate systems  throughout the calculations.  \\
\textit{Proper frame:}
\be
 ds^2 = e^{-2A(y)}(-h(y) d\tau^2 + d\x^2) + \frac{dy^2}{h(y)} \,.
\ee
\\
\textit{Conformal frame:}
\be
ds^2 = \omega(z)^2 \left(-f(z) d\tau^2 + d\x^2 + \frac{dz^2}{f(z)} \right) \,.
\ee
\\
\textit{Brane cosmology frame:}
\be
ds^2 = -n(r)^2 d\tau^2 + \frac{r^2}{\ell^2} d\x^2 + b(r)^2 dr^2\,.
\ee
The relations between them depend on the integration constants; they are specified along the calculations. 

\subsection{Superpotential}

\label{se:W}

When solving the scalar-gravity system it is convenient to introduce the superpotential function $W(\phi)$ satisfying (see 
\textit{e.g.} \cite{Cabrer:2009we,Megias:2019vdb})
\be
 V(\bar\phi) = \frac{1}{6 M_5^3} \left( \frac{1}{4} \left( \frac{\partial W}{\partial \bar\phi} \right)^2 -  W(\bar\phi)^2  \right) \label{eq:W}\,.
\ee
We find that some relations in the various backgrounds considered  (\textit{i.e.} AdSS, LD, ALD)  can be expressed in a unified fashion in terms of the superpotential. 
We introduce the reduced superpotential evaluated on the brane, $\bar W _b \equiv\frac{W(\bar v_{\brane})}  {3M_5^3}\,. $

First, we find that the relation between $M_5$ and $M_{\rm Pl}$ is generally given by 
\be M_5^3 = \frac{1}{2} \bar W_b M_{\rm Pl}^2 \,.
\ee
This is verified for \textit{e.g.}~Eqs.~(\ref{eq:M5M4_LD}) and (\ref{eq:M5M4_LDA}). As a result the low-energy condition Eq.\,\eqref{eq:smallT} can be expressed as 
\be 
\frac{1}{M^2_{\rm Pl}}|T^b_{\mu\nu}|\ll \bar W_b^2 \label{eq:smallTW}\,. 
\ee

Second, we find that the effective 4D cosmological constant is expressed as 
\begin{equation}
\Lambda_4 = \frac{1}{2M_5^3} \left( -\frac{3}{2} M_5^3 \bar W_b^2 + \frac{\Lambda_{\brane}^2}{6 M_5^3}\right)  \,.
\end{equation}
The tuned value $\Lambda_4 = 0$ is then obtained for
\begin{equation}
\Lambda_{\brane} =  3 M_5^3 \bar W_b \,. \label{eq:Lambdab_gen}
\end{equation}
With this tuning we have, for example, 
$\rho^\Lambda =  \frac{3}{2} M_5^3 \bar W_b$.

\subsection{AdS-Schwarzschild}

In the conformal frame the solution of the equations of motion is given by
\begin{equation}
    f_{\AdSS}(z) = C_A^2 \left( 1 - \frac{(z-z_\ast)^4}{(z_h - z_\ast)^4} \right) \,, \qquad \omega_{\AdSS}(z) = C_A \frac{\ell}{z - z_\ast} \,, \qquad \bar\phi_{\AdSS}(z) = C_\phi \,,
\end{equation}
where $C_A$, $C_\phi$, $z_\ast$, and the horizon position $z_h$ are integration constants. In the brane cosmology frame one finds
\begin{eqnarray}
n_{\AdSS}(r) =  C_A \frac{r}{\ell} \sqrt{1 - \frac{r_h^4}{r^4}} \,, \qquad 
b_{\AdSS}(r) = \frac{\ell}{r} \frac{1}{\sqrt{1 - \frac{r_h^4}{r^4}}} \,, \qquad \bar\phi_{\AdSS}(r) = C_\phi \,,
\end{eqnarray}
and the relation between both coordinates is given by $\omega_{\AdSS}(z) = \frac{r}{\ell}$. We can already notice that $z_\ast$ is an irrelevant  shift symmetry in the coordinate $z$, which does not have counterpart in the brane cosmology frame. We can thus set $z_\ast = 0$ without loss of generality. The $C_\phi$ constant is also physically irrelevant because $V(\phi) = $ cte.

The temperature and Weyl energy turn out to be 
\begin{equation}
T_h = C_A \frac{r_h}{\pi \ell^2} \,, \qquad \rho^W_{\AdSS}(r_{\brane}) = \frac{3 M_{\rm Pl}^2}{\ell^2} \frac{r_h^4}{r_{\brane}^4} \,. 
\end{equation}
The $C_A$ constant affects the horizon temperature but not the Weyl energy. 
It can be eliminated by a constant rescaling of the coordinates, so that we can set $C_A=1$, which gives the usual AdS-Schwarzschild metric of section \ref{se:AdSS}. The  Weyl energy depends on $r_h$, the horizon position. This is the only physically meaningful integration constant.

\subsection{Linear dilaton}
\label{sucsec:app_LD}

\subsubsection*{Proper frame}
In the LD model, the solution of the equations of motion in proper coordinates is written as 
\begin{eqnarray}
h_{\LD}(y) &=& C_A^2 \left( 1 - e^{3 (A_{\LD}(y) - A_{\LD}(y_h))}\right) = C_A^2 \left( 1- \left(\frac{1 - y_h/y_s}{1-y/y_s} \right)^3\right)  \,, \\ 
A_{\LD}(y) &=& -\log\left( 1 - \frac{y}{y_s} \right) \,, \\ 
\bar \phi_{\LD}(y) &=& -\log\left(1 - \frac{y}{y_s} \right) + \log(C_A C_S \ell) \,,     
\end{eqnarray}
where $C_S=1/y_s$, with $C_A$, $y_s$, and $y_h$ integration constants. In this solution we have neglected an irrelevant shift symmetry in the coordinate $y$. 

\subsubsection*{Conformal frame}

In the conformal frame the solutions are
\begin{eqnarray}
f_{\LD}(z) &=& C_A^2 \left(  1 - e^{3(A_{\LD}(z) - A_{\LD}(z_h))}\right) = C_A^2 \left( 1 - e^{3 C_S (z - z_h) } \right)  \,, \\
\omega_{\LD}(z) &=& e^{-A_{\LD}(z)}= e^{-C_S z} \,, \\
\bar \phi_{\LD}(z) &=& C_S z + \log(C_A C_S \ell) \,,     
\end{eqnarray}
where $C_A$, $C_S$, and $z_h$ are integration constants. As in the AdSS case, we have neglected an irrelevant shift symmetry in $z$. The effective Schr\"odinger potential for the graviton is given by
\begin{equation}
V_{\bar\eta}(z) = \frac{9}{4} A_{\LD}^\prime(z)^2 - \frac{3}{2} A_{\LD}^{\prime\prime}(z) = \left( \frac{3}{2} C_S\right)^2 \equiv \sigma^2 \,.
\end{equation}
The $C_S$ constant has a physical meaning: it is
identified with the  $\bar\eta$ scale (see main text) which is proportional to the mass gap in the spectrum (see also e.g.~\cite{Antoniadis:2011qw,Cox:2012ee,Megias:2021mgj} for  discussions). 

\subsubsection*{Brane cosmology frame}

In the brane cosmology frame we find
\begin{eqnarray}
n_{\LD}(r) =  C_A \frac{r}{\ell} \sqrt{1 - \frac{r_h^3}{r^3}} \,, \quad 
b_{\LD}(r) = C_B \frac{1}{\sqrt{1 - \frac{r_h^3}{r^3}}} \,, \quad \bar\phi_{\LD}(r) = -\log\left( C_B \frac{r}{\ell} \right) \,.
\end{eqnarray}
In these coordinates the solution involves three integration constants: $C_A$, $C_B$ and $r_h$. The gap is computed as
\begin{equation}
V_{\bar\eta}(r) = \frac{3}{2} \frac{r^2}{\ell^4} \frac{1}{n(r)^2 b(r)^2} \left[ \frac{5}{2} - r \left( \frac{n^\prime(r)}{n(r)} + \frac{b^\prime(r)}{b(r)} \right) \right]  = \sigma^2 = \left( \frac{3}{2} \bar\eta\right)^2 \,, \label{eq:V_eta}
\end{equation}
with
\begin{equation}
\bar\eta = \frac{1}{C_A C_B \ell} \,.
\end{equation}
The relation between $y$ and $r$ coordinates in the LD model, not assuming any particular value for the integration constants, is
\begin{equation}
\frac{r}{\ell} = 1 - \frac{y}{y_s} \,, \label{eq:dydr1}
\end{equation}
and
\begin{equation}
\left( \frac{\ell}{r} n_{\LD}(r) \right)^2 =  h(y)\,, \qquad \frac{r}{\ell} = e^{-A_{\LD}(y)}  \,, \qquad \frac{dy}{dr} = - C_A C_B \,.  \label{eq:yr_LD}
\end{equation} 
Consistency of Eqs.~(\ref{eq:dydr1}) and (\ref{eq:yr_LD}) requires $C_B = y_s / (C_A \ell)$. The relation between $z$ and $r$ coordinates is $e^{-C_S z} = \frac{r}{\ell}$.

\subsubsection*{General result}

Finally, the temperature and Weyl energy turn out to be 
\begin{equation}
T_h = \frac{C_A}{C_B} \frac{3}{4\pi}  \frac{1}{\ell} \,, \qquad \rho^{W}_{\LD}(r_{\brane}) = \frac{9}{4} \frac{M_{\rm Pl}^2}{C_B^2} \frac{r_h^3}{r_{\brane}^5} \,.  \label{eq:Th_C_LD}
\end{equation}
The $C_A$ constant affects the horizon temperature but not the Weyl energy. Unlike the AdSS case, $\rho^W_{\LD}(r_{\brane})$ depends on both the horizon position and another integration constant,~$C_B$. This constant corresponds to the boundary value of $b(r)$, {\it i.e.}  $C_B = \lim_{r\to \infty} b(r)$.
Moreover, $C_B$ contributes to the scalar vev via $-\log C_B$. 

\subsubsection*{Fixing the constants}

In the present work we have considered the hypothesis that the scalar vev at the brane $r = r_{\brane}$ is constant, {\it i.e.}~it does not evolve with time. As a result, we should set
\begin{equation}
C_B = e^{-\bar v_b} \frac{\ell}{r_{\brane}} \,, \label{eq:CB_LD1}
\end{equation}
thus implying 
\begin{equation}
\bar\phi_{\LD}(r) = \bar v_{\brane} - \log\left( \frac{r}{r_{\brane}}\right) \,,
\end{equation}
and $\bar\phi_{\LD}(r_{\brane}) = \bar v_{\brane}$. As in the AdSS metric, $C_A = 1$ leads to the zero temperature solution near the boundary, {\it i.e.}~far from the black hole horizon.
 In summary, we can set
\begin{equation}
    C_A = 1 \,, \qquad  C_S = \eta \,,\qquad C_B = \frac{1}{C_A C_S \ell} = \frac{1}{\bar\eta \ell} \,, \label{eq:CB_LD2}
\end{equation}
which leads to the solution of section~\ref{subsec:LD}. Combination of Eqs.~(\ref{eq:CB_LD1}) and (\ref{eq:CB_LD2}) leads to
\begin{equation}
\bar\eta =  \eta \frac{r_{\brane}}{\ell} \,, \qquad \textrm{where} \qquad \eta \equiv \frac{e^{\bar v_{\brane}}}{\ell}\,. \label{eq:eta_etabar_LD}
\end{equation}
Then, the temperature and Weyl energy can be written as 
\begin{equation}
T_h = \frac{3}{4\pi} \bar\eta \,, \qquad \mathcal \rho^W_{\LD}(r_{\brane}) = \frac{9}{4} \eta^2 M_{\rm Pl}^2 \frac{r_h^3}{r_{\brane}^3} \,. \label{eq:Th_rhoW_LD}
\end{equation}
%

\subsection{Linear dilaton with AdS asymptotics}
\label{subsec:app_ALD}

\subsubsection*{Proper frame}

In the ALD model, the solution in proper coordinates  is
\begin{eqnarray}
A_\ALD(y) &=& C_{a} \frac{y}{\ell} - \log\left( 1 - \frac{y}{y_s}\right) \,, \\  h_\ALD(y) &=& \frac{1}{C_{a}^2} \left( 1 - \frac{ \int_{-\infty}^y d{\bar y} \, e^{4A_{\ALD}({\bar y})} }{ \int_{-\infty}^{y_h} d{\bar y} \, e^{4A_{\ALD}({\bar y})} } \right) \,, \\
\bar\phi_\ALD(y) &=& -\log\left( \frac{y_s - y}{\ell} \right) - \log C_{a} \,,
\end{eqnarray}
with $C_{a}$, $y_s$ and $y_h$ as integration constants. As in the previous models, we have neglected an irrelevant shift symmetry in $y$.

\subsubsection*{Conformal frame}

The relation between $y$ and $z$ coordinates is
\begin{equation}
\frac{dy}{dz} = - e^{-A_{\ALD}(y)} = - e^{-C_a y/\ell} \left( 1 - \frac{y}{y_s} \right) \,.
\end{equation}
Then, the effective Schr\"odinger potential for the graviton is given by
\begin{eqnarray}
V_{\bar\eta} &=& \frac{9}{4} A_{\ALD}^\prime(z)^2 - \frac{3}{2} A_{\ALD}^{\prime\prime}(z) = \frac{3}{4} e^{-2A(y)} \left( 5 A_{\ALD}^\prime(y)^2 - 2 A_{\ALD}^{\prime\prime}(y) \right) \nonumber \\
&=& \frac{3}{4} \frac{1}{y_s^2} e^{-2 C_a y/\ell} \left( 3 + 10 C_a \left( 1 - \frac{y}{y_s}\right) \frac{y_s}{\ell}   + 5 C_a^2 \left(1 - \frac{y}{y_s} \right)^2 \frac{y_s^2}{\ell^2}\right)   \,.
\end{eqnarray}
The gap is then,
\begin{equation}
\sigma^2 = \lim_{z\to +\infty} V_{\bar\eta}(z) = \lim_{y \to y_s} V_{\bar\eta}(y) = \left( \frac{3}{2} \bar\eta \right)^2 \qquad \textrm{with} \qquad \bar\eta =  \frac{1}{y_s} e^{- C_a y_s/\ell} \,. \label{eq:eta_ALD}
\end{equation}

\subsubsection*{Brane cosmology frame}

In the brane cosmology frame we find
\begin{eqnarray}
n_\ALD(r) &=& \frac{1}{C_a} \frac{r}{\ell} \sqrt{\bar h_{\ALD}(y(r))} \,,  \label{eq:n_ALD} \\  
b_\ALD(r) &=&  \frac{1}{\sqrt{\bar h_{\ALD}(y(r))}} \frac{\ell}{r} \frac{\mathcal W\left( C_{B} \frac{r}{\ell} \right) }{1 + \mathcal W\left( C_{B} \frac{r}{\ell} \right) }  \,, \label{eq:b_ALD} \\
\bar\phi_\ALD(r) &=& -\log \mathcal W\left( C_{B} \frac{r}{\ell} \right) \,, \label{eq:phi_ALD}
\end{eqnarray}
where we have defined
\begin{equation}
\bar h_\ALD(y) =  1 - \frac{ \int_{-\infty}^y d{\bar y} \, e^{4A_{\ALD}({\bar y})} }{ \int_{-\infty}^{y_h} d{\bar y} \, e^{4A_{\ALD}({\bar y})} }  \,.
\end{equation}
The relation between $r$ and $y$ coordinates is
\begin{equation}
\frac{r}{\ell} = e^{-A_{\ALD}(y)} = \left( 1 - \frac{y}{y_s} \right)  e^{-C_{a} y/\ell}  \,,
\end{equation}
The full domain in proper coordinates is $y\in (-\infty,y_s]$ while the domain in the cosmological coordinates is $r\in[0,\infty)$. The brane position at present times is chosen by convention to be $a(t_0)=1$, \textit{i.e.}~$r=\ell$, or $y=0$. 

The integration constants in the brane cosmology coordinates are $C_{a}$, $C_B$ and $r_h \equiv r(y_h)$, and their relation with $\bar\eta$ is $C_B = C_a/(\bar\eta \ell)$. This relation follows from the gap, which is computed from the effective Schr\"odinger potential for the graviton (cf.~Eq.~(\ref{eq:V_eta}))
\begin{equation}
V_{\bar\eta}(r) = \frac{3}{4} C_a^2 \frac{r^2}{\ell^4} \frac{3 + 10 \mathcal W\left( C_{B} \frac{r}{\ell} \right) + 5 \mathcal W^2\left( C_{B} \frac{r}{\ell} \right) }{\mathcal W^2\left( C_{B} \frac{r}{\ell} \right)}  \,,
\end{equation}
as
\begin{equation}
\sigma^2 = \lim_{r\to 0} V_{\bar\eta}(r) = \left( \frac{3}{2} \bar\eta \right)^2 \qquad \textrm{with} \qquad \bar\eta = \frac{C_a}{C_B \ell} \,.  \label{eq:eta_ALD_r}
\end{equation}

\subsubsection*{General result}

Finally, the temperature and Weyl energy turn out to be
\begin{equation}
T_h = \frac{1}{C_{a}^2} \frac{\bar\eta}{4\pi \chi_h^3 I(\chi_h)} \,, \qquad \rho^W_{\ALD}(r_{\brane}) = \frac{3 M_{\rm Pl}^2}{4 \ell^2} \frac{1}{\chi_b^4 I(\chi_h)}  \left( 1 + \frac{1}{\mathcal W(\chi_{\brane})} \right)   \,, \label{eq:Th_Cr_ALD}
\end{equation}
where
\begin{equation}
  I(\chi) \equiv \int_{\chi}^\infty \frac{dx}{x^5} \frac{\mathcal W(x)}{1 + \mathcal W(x)} =  \frac{1}{3\chi^4} \left[ \mathcal W(\chi) - 2 \mathcal W^2(\chi) + 8 \mathcal W^3(\chi) + 32 \chi^4 {\rm Ei}(-4 \mathcal W(\chi))  \right]  \,, \label{eq:I_chi}
\end{equation}
and
\begin{equation}
\chi_{\brane} \equiv  C_{a} \frac{r_{\brane}}{\bar\eta \ell^2}  \,, \qquad  \chi_h \equiv  C_{a} \frac{r_h}{\bar\eta \ell^2} \,,
\end{equation}
while ${\rm Ei}(z)$ is the exponential integral function.  

\subsubsection*{Fixing the constants}

As mentioned in Sec.~\ref{sucsec:app_LD}, we  fix the integration constants in the present work in such a way that the vev of the scalar field at the brane is constant. According to Eq.~(\ref{eq:phi_ALD}), this implies to assume $C_B \propto 1/r_{\brane}$, {\it i.e.}
\begin{equation}
C_B = c \frac{\ell}{r_{\brane}} \,, \label{eq:CB_ALD}
\end{equation}
where $c$ is a dimensionless constant. Then, the vev at the brane is
\begin{equation}
\bar v_{\brane} \equiv \bar\phi_{\ALD}(r_{\brane}) = -\log \mathcal W(c) \,.
\end{equation}
By solving this equation for $c$, one has that this parameter is related to the vev as
\begin{equation}
c = e^{ -\bar v_{\brane}  + e^{-\bar v_{\brane}}  } \,.
\end{equation}
The other integration constants turn out to be related to each other by~\footnote{An alternative expression for $C_a$ can be obtained from Eqs.~(\ref{eq:eta_ALD}), (\ref{eq:eta_ALD_r}) and (\ref{eq:CB_ALD}), which yields 
\begin{equation}
C_a = \frac{\ell}{y_s} \mathcal W\left(c \ell/r_{\brane} \right) \,,
\end{equation}
so that the function $r(y)$ and its inverse, {\it i.e.} $y=y(r)$, are given by Eq.~(\ref{eq:ry_inv}).}
\begin{equation}
C_a = \bar\eta \ell C_B = c \frac{\bar\eta \ell^2}{r_{\brane}} \,. \label{eq:CA_Ca_CB}
\end{equation}
Then, one finds
\begin{equation}
\rho^W_{\ALD}(r_{\brane}) = \frac{3 M_{\rm Pl}^2}{4\ell^2} \frac{\left( 1 + e^{\bar v_{\brane}} \right) }{c^4} \frac{1}{I(\chi_h)}\,, 
\end{equation}
where
\begin{equation}
\chi_{\brane} = c \,, \qquad \chi_h = c \frac{r_h}{r_{\brane}}  \,.
\end{equation}
Notice that $r_{\brane}$ small (large) corresponds to $\chi_h$ large (small). 

\subsubsection*{Asymptotics}

By using the asymptotic behaviors of the Lambert function, 
\begin{equation}
\mathcal W(\chi)  \simeq \left\{
    \begin{array}{cc}
\chi   & \qquad \chi \ll 1  \\
\log \chi  &  \qquad \chi \gg 1
\end{array} \,, \right. \label{eq:W}
\end{equation}
one finds
\begin{equation}
I(\chi)  \simeq \left\{
    \begin{array}{cc}
\frac{1}{3\chi^3}   & \qquad \chi \ll 1  \\
\frac{1}{4\chi^4}  &  \qquad \chi \gg 1
\end{array} \,. \right. \label{eq:Iasymptotic}
\end{equation}
From this one obtains in the regime $c \ll 1$ the same results for the temperature and Weyl energy as in the LD background, cf.~Eqs.~(\ref{eq:Th_C_LD}), (\ref{eq:CB_LD2})-(\ref{eq:Th_rhoW_LD}),  while in the regime $c \gg 1$ the following behaviors are obtained
\begin{equation}
  T_h \simeq \left\{
    \begin{array}{cc}
\frac{1}{C_a^2} \frac{3\bar \eta}{4\pi}  &   \,\;\quad\quad\;\;  {\rm if} ~~  1 \ll c < \ell/r_{h,0} \\
\frac{1}{C_a^2} \frac{c\bar\eta}{\pi} \frac{r_h(r_b)}{r_b} \simeq \frac{1}{C_a^2} \frac{c \bar\eta}{\pi} \frac{r_{h,0}}{r_b}    &  \,\;\quad\quad\;\;  {\rm if} ~~  1 \ll \ell/r_{h,0} < c
\end{array} \,, \right. \label{eq:Th}
\end{equation}
for the temperature, and
\begin{equation}
\rho^W_{\ALD}(r_{\brane}) \simeq   \left\{
    \begin{array}{cc}
\frac{9}{4\ell^2}  \frac{1}{c}  M_{\rm Pl}^2 \frac{r_h^3(r_b)}{r^3_{\brane}} \simeq \frac{9}{4\ell} \frac{1}{c} M_{\rm Pl}^2  \frac{r_{h,0}^3}{r_{\brane}^4}    &   \,\;\quad\quad\;\;  {\rm if} ~~  1 \ll c < \ell/r_{h,0} \\
\frac{3}{\ell^2}  M_{\rm Pl}^2 \frac{r_h^4(r_b)}{r_{\brane}^4} \simeq \frac{3}{\ell^2}  M_{\rm Pl}^2 \frac{r_{h,0}^4}{r_{\brane}^4}   &  \,\;\quad\quad\;\;  {\rm if} ~~  1 \ll \ell/r_{h,0} < c
\end{array} \,, \right. \label{eq:C} 
\end{equation}
for the Weyl energy. In the second equalities of these formulas we have used that the solution of the conservation equation, Eq.~(\ref{eq:5Dcont}), demands {that the black hole horizon depends on the brane position}, \textit{i.e.}~$r_h = r_h(r_b)$. This dependence turns out to be almost constant for $c \ll 1$ and $1 \ll \ell/r_{h,0} < c$, while it behaves as $r_h(r_b) \simeq r_{h,0} \left( \ell/r_b\right)^{1/3}$ for $1 \ll c < \ell/r_{h,0}$ and $r_{\brane,\ast} < r_b$, see discussion around Eq.~(\ref{eq:rhrb_analytic}).

In summary, we find that while the behavior of $\rho^W_{\ALD}(r_{\brane})$ is $\propto r_b^{-3}$ for small values of $c$, it changes to $\propto r_b^{-4}$ at large values of $c$, signaling the change from a 4D cold matter regime (with $w = 0$), to a 4D radiation regime (with $w = \frac{1}{3}$). The same conclusion is obtained for the contributions $\rho^\phi_{\ALD}(r_b) + \rho^\Lambda_{\ALD}$, although the latter becomes subdominant with respect to $\rho^W_{\ALD}(r_b)$ in the regime $c \gg 1$. 

For completeness, we provide the explicit exact expression of $\rho_{\rm eff,\ALD}(r_{\brane}) = \rho_{\ALD}^W(r_{\brane}) + \rho^\phi_{\ALD}(r_{\brane}) + \rho^\Lambda_{\ALD}$. This is given by~\footnote{The expression of $\rho_{\rm eff,\ALD}$ as a function of $r_{\brane}$ and $r_{h}$ is given by
\begin{equation}
\rho_{\rm eff,\ALD}(r_{\brane},r_h) = \frac{3}{4} \frac{M_{\rm Pl}^2}{\ell^2} \frac{\mathcal W(c) + \mathcal W^2(c) + c^4 I(c)}{c^4 \mathcal W^2(c) I(c\, r_h / r_{\brane})} \,. \label{eq:rhoeff_rhrb}
\end{equation}
From a comparison with Eq.~(\ref{eq:rhoeff_ALD_analytic}), it turns out that the dependence $r_h = r_h(r_{\brane})$ is obtained from the solution of the equation $I(c\, r_h(r_{\brane})/r_{\brane}) = I(c\, r_{h,0}/\ell)  \left( r_{\brane}/\ell \right)^{3(1+w_{\rm eff}(c))}$. Alternatively, $r_h(r_{\brane})$ can be obtained by plugging Eq.~(\ref{eq:rhoeff_rhrb}) into the conservation equation~(\ref{eq:5Dcont}), and solving the resulting first order differential equation for $r_h(r_{\brane})$ with boundary condition $r_h(\ell) = r_{h,0}$. We have checked that both methods lead to the same result.}
\begin{equation}
\rho_{\rm eff,\ALD}(r_{\brane}) = \rho_{\rm eff,0} \left( \frac{\ell}{r_{\brane}}\right)^{3(1 + w_{\rm eff}(c))} \,, \label{eq:rhoeff_ALD_analytic}
\end{equation}
where
\begin{equation}
\rho_{\rm eff,0} = \frac{3}{4} \frac{M_{\rm Pl}^2}{\ell^2} \frac{\mathcal W(c) + \mathcal W^2(c) + c^4 I(c)}{c^4 \mathcal W^2(c) I(c\, r_{h,0}/\ell)} \label{eq:rhoeff0_ALD_analytic}
\end{equation}
is the effective energy density at present times $t = t_0$, and
\begin{equation}
w_{\rm eff}(c) \equiv  \frac{P_{\rm eff,\ALD}}{\rho_{\rm eff,\ALD}} = \frac{1}{3} \frac{ \mathcal W(c) + \mathcal W^2(c) - 3 c^4 I(c)}{\mathcal W(c) + \mathcal W^2(c) + c^4 I(c)} \label{eq:weff_analytic}
\end{equation}
is the equation-of-state parameter.

Although it is not relevant for the result of $\rho_{\rm eff,\ALD}(r_b)$, it is natural to fix $C_a=1$ as this allows to connect with the zero temperature solution near the boundary, cf.~Eq.~(\ref{eq:n_ALD}). By using Eq.~(\ref{eq:CA_Ca_CB}), this leads to
\begin{equation}
\bar \eta = \eta \frac{r_b}{\ell} \,,  \qquad \textrm{where} \qquad  \eta \equiv 
 \frac{1}{c\ell} =\frac{e^{\bar v_{\brane} - e^{-\bar v_{\brane}}}}{\ell} \,, \label{eq:eta_etabar_ALD}
\end{equation}
which is the  counterpart in the ALD model of Eq.~(\ref{eq:eta_etabar_LD}). Notice that $\eta$, as defined in Eq.~(\ref{eq:eta_etabar_ALD}) for the ALD model, tends to the corresponding value of $\eta$ in the LD model, given in Eq.~(\ref{eq:eta_etabar_LD}), when considering the limit $\bar v_b \gg 1$. Moreover, in the opposite limit where $c\to\infty$, \textit{i.e.}~$\bar v_b<0$ and $|\bar v_b|\gg 1$, the parameter $\eta$ in the ALD model tends to zero, as expected since the AdSS spectrum is gapless.

\section{Brane kinematics and  conservation law }

\label{app:5D_continuity_eq}

The conservation equation of $\rho_{\rm eff}$ is given in Eq.\,\eqref{eq:5Dcont} and reproduced here,
\begin{equation}
\dot \rho_{\rm eff} + 4 H \rho_{\rm eff} + H T^{{\rm eff}\,\mu}_\mu    = - 2 \left( 1 + \frac{\rho_{\brane}}{\Lambda_{\brane}}\right) T^\phi_{MN} u^M n^N   \,, \label{eq:5Dcont_App}
\end{equation}
 Here we evaluate $T^\phi_{MN} u^Mn^N$ with the metric
\be
ds^2= -n^2 d\tau^2+\frac{r^2}{\ell^2}d{\x}^2 +b^2 dr^2 
\ee
in the linear dilaton and  asymptotically linear dilaton backgrounds. An AdS version of this analysis can be found in \cite{Tanaka:2003eg,Langlois:2003zb}.

The trajectory of the brane can be represented by functions $\tau(t)$, $r_b(t)$ where $t$ is the brane proper time.  
$u^M$ is the timelike unit  vector for brane observers, $u^M= (\dot \tau,{\bm 0}  , \dot r_b)$, where the dot represents $\partial/\partial t$. The Hubble scale is related to $r_b$ by  $\dot r_b = H r_b$. 
The  normalization $u^Mu_M=-1$ implies $\dot \tau = \frac{\sqrt{1+\dot r_b^2 b^2}}{n} $ thus 
\be
u^M= \left( \frac{\sqrt{1+\dot r_b^2 b^2}}{n},{\bm 0}  , \dot r_b\right) \,.
\ee
The  unit  vector normal to the brane $n^M$ satisfies $n^M u_M =0 $  and $n^M n_M=1$, we find 
\be
n^M= \left(\dot r_b \frac{b}{n},  {\bm 0} , \frac{\sqrt{1+\dot r_b^2 b^2}}{b}   \right) \,.
\ee

We then evaluate the stress tensor in the linear dilaton background. Using that $\frac{\partial r_b}{\partial \tau}=n \dot r_b $ and the solution $\bar \phi=\bar v_b -\log(r/r_b)$ we get 
\begin{eqnarray}
\frac{1}{3M_5^3}T^\phi_{00} &=& \frac{1}{2}(\partial_\tau \bar\phi)^2+\frac{n^2}{2b^2}(\partial_r \bar\phi)^2|_{r=r_b} +n^2 \bar V = \frac{1}{2}n^2 H^2+\frac{n^2}{2b^2r^2_b}  +n^2 \bar V \,, \\
\frac{1}{3M_5^3} T^\phi_{05} &=& \frac{1}{3M_5^3} T_{50} = (\partial_\tau \bar\phi) (\partial_r \bar\phi)|_{r = r_{\brane}}  =  -\frac{nH}{r_b} \,, \\
\frac{1}{3M_5^3} T^\phi_{55} &=& \frac{1}{2}(\partial_r  \bar\phi)^2|_{r=r_b}+\frac{b^2}{2n^2}(\partial_\tau \bar\phi)^2  -b^2 \bar V =  \frac{1}{2r^2_b}+\frac{b^2}{2} H^2 -b^2 \bar V\,.
\end{eqnarray}
The result is
\begin{eqnarray}
T^\phi_{MN} u^M n^N &=& 3 M_5^3 \frac{H}{r_b b} \left[ -H^2 r_b^2 b^2  + (1 +  H^2 r_b^2 b^2 ) \left( \sqrt{1 +  H^2 r_b^2 b^2 } - 1 \right)  \right] \,. 
\end{eqnarray}
In the low-energy regime $H\ll \eta$ we then find
\be
T^\phi_{MN}u^Mn^N = 0+  O\left(\frac{H^3}{\eta^3}\right) \,. \label{eq:Tun}
\ee
The fact that the leading term vanishes  is nontrivial. As a result,  the $T^\phi_{MN}u^Mn^N$  term is negligible in the conservation law. The same cancellation as in Eq.~(\ref{eq:Tun}) is obtained in the ALD background.

Finally, for completeness we also verify the general relation of Eq.~(\ref{eq:modified-japanese}) 
in the low-energy regime. We have
\begin{align}
T^\phi_{MN} n^M n^N &=  \frac{3}{2} M_5^3 \left[ \frac{1}{r_b^2 b^2} + 2 H^4 r_b^2 b^2 + H^2 \left( 3 - 4 \sqrt{1 + H^2 r_b^2 b^2 } \right) - 2 \bar V\right]  \nn
\\
&=  3 M_5^3 \left( \frac{1}{2 r_b^2 b^2} - \bar V\right) + O\left(\frac{H^2}{\eta^2}\right) \,, 
\label{eq:Tnn}
\end{align}
 and thus 
\be
 2  \frac{M^2_{\rm Pl}}{M_5^3}H T^\phi_{MN} n^M n^M + H \tau_{\,\, \mu}^{{\rm \Lambda}\, \mu} 
= 6 M_{\rm Pl}^2 H \left(\frac{1}{2} \frac{\bar \phi^\prime(r_b)^2}{b^2} - \bar V \right)  + H \tau_{\,\, \mu}^{{\rm \Lambda}\, \mu}  = H  T_{\,\, \mu}^{{\rm eff}\, \mu} \,.  \label{eq:Tnn_2}
\ee
 We perform the same calculations in the ALD model and get the same result as (\ref{eq:Tnn_2}).

\bibliographystyle{JHEP}
\bibliography{biblio}

\end{document}